\documentclass[10pt,journal,compsoc]{IEEEtran}
\usepackage{amsmath,amsfonts}
\usepackage{algorithmic}
\usepackage{algorithm}
\usepackage{array}
\usepackage[caption=false,font=normalsize,labelfont=sf,textfont=sf]{subfig}
\usepackage{textcomp}
\usepackage{stfloats}
\usepackage{url}
\usepackage{verbatim}
\usepackage{graphicx}
\usepackage{cite}
\usepackage{color}
\definecolor{AAA}{rgb}{1.0, 0.13, 0.32}
\definecolor{BBB}{rgb}{0.2, 0.1, 1}

\usepackage{booktabs} % For formal tables
\usepackage{amssymb}
\usepackage[numbers,sort]{natbib}
\usepackage{lineno}
\usepackage{soul}

\usepackage[percent]{overpic} %?
\usepackage{stfloats}
\usepackage{wrapfig}
 \usepackage{multirow}
\usepackage{svg} % 需使用包\usepackage[acronym]{glossaries}

\begin{document}
% \linenumbers
\title{Towards Voronoi Diagrams of Surface Patches}


\author{Pengfei Wang, Jiantao Song, Lei Wang, Shiqing Xin, Dong-Ming Yan, Shuangmin Chen, Changhe Tu, Wenping Wang
% \IEEEcompsocitemizethanks{
% \IEEEcompsocthanksitem P. Wang, J. Song, L. Wang, S. Xin, C. Tu are with the School of Computer Science, Shandong University. S. Xin is the corresponding author (email: xinshiqing@sdu.edu.cn)
% \IEEEcompsocthanksitem D.-M. Yan is with MAIS, Institute of Automation, Chinese Academy of Sciences, Beijing 100190, China, and the School of Artificial Intelligence, University of Chinese Academy of Sciences, Beijing 100049, China. E-mail:  yandongming@gmail.com.
% \IEEEcompsocthanksitem S. Chen is with the School of Information and Technology, Qingdao University of Science and Technology.
% \IEEEcompsocthanksitem W. Wang is with the Computer Science and Engineering, Texas A\&M University
% }
}

\maketitle



\begin{abstract}
Extraction of a high-fidelity 3D medial axis is a crucial operation in CAD. When dealing with a polygonal model as input, ensuring accuracy and tidiness becomes challenging due to discretization errors inherent in the mesh surface. Commonly, existing approaches yield medial-axis surfaces with various artifacts, including zigzag boundaries, bumpy surfaces, unwanted spikes, and non-smooth stitching curves.
Considering that the surface of a CAD model can be easily decomposed into a collection of surface patches, its 3D medial axis can be extracted by computing the Voronoi diagram of these surface patches, where each surface patch serves as a generator. However, no solver currently exists for accurately computing such an extended Voronoi diagram.
Under the assumption that each generator defines a linear distance field over a sufficiently small range, our approach operates by tetrahedralizing the region of interest and computing the medial axis within each tetrahedral element. Just as SurfaceVoronoi computes surface-based Voronoi diagrams by cutting a 3D prism with 3D planes (each plane encodes a linear field in a triangle), the key operation in this paper is to conduct the hyperplane cutting process in 4D, where each hyperplane encodes a linear field in a tetrahedron.
In comparison with the state-of-the-art, our algorithm produces better outcomes. Furthermore, it can also be used to compute the offset surface.
\end{abstract}


\begin{figure*}[htb]
	\centering
\begin{overpic}
[width=.98\linewidth]{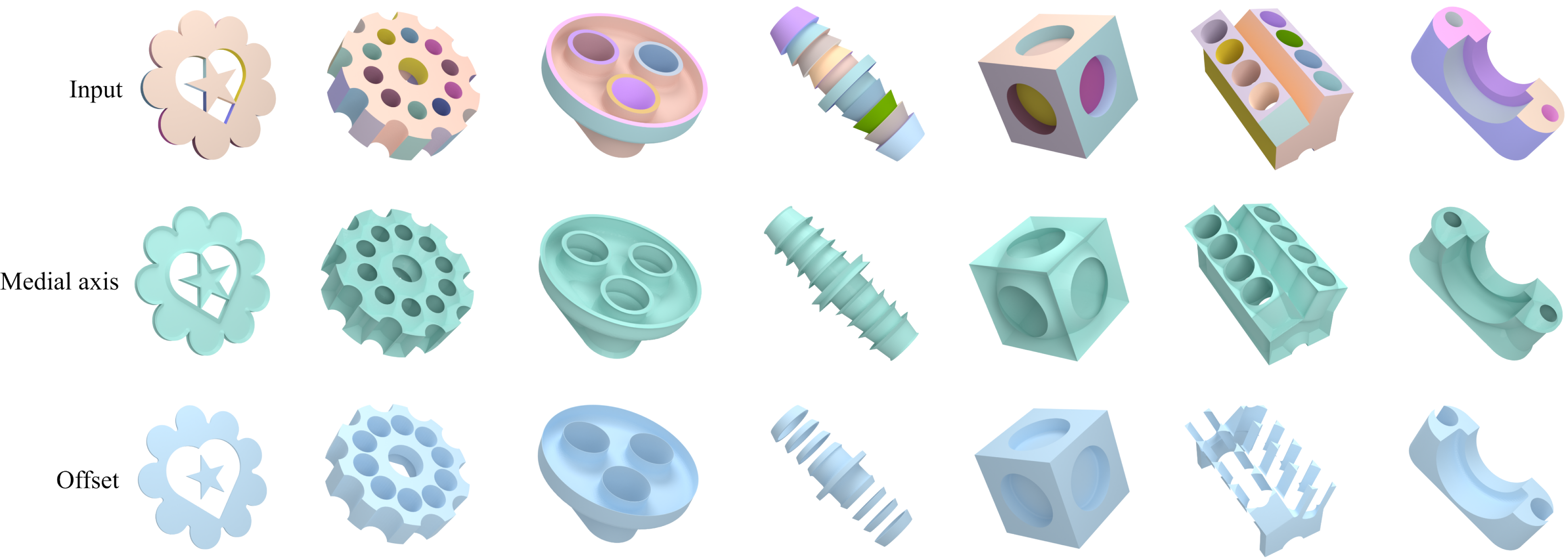}
\end{overpic}
\caption{
This paper suggests computing the medial axis of CAD models via Voronoi diagrams of surface patches, treating each patch as an individual computational unit. Interestingly, this computational technique can also be used to calculate offsets.
Top: Input polygonal models (surface patches are visualized in a color-coded style).
Middle: Medial-axis surfaces.
Bottom: Inward offset surfaces.
}
\label{FIG:teaser}
\end{figure*}

\section{Introduction}
Voronoi diagrams serve the purpose of partitioning a given space into subregions based on proximity. Beyond their direct applications in proximity queries and collision detection~\cite{mirtich1998v,cuenca2022collision}, Voronoi diagrams find utility in a diverse range of fields, including surface reconstruction~\cite{boltcheva2017surface}, robot motion planning~\cite{garrido2015mobile}, non-photorealistic rendering~\cite{lindemeier2015hardware}, surface simplification~\cite{xin2011isotropic}, mesh generation~\cite{yan2013efficient}, and shape analysis~\cite{ying2015point}, among others.

Voronoi diagrams showcase numerous variants depending on specific domains, metrics, and generator types, with the most commonly employed version defined in Euclidean spaces using point generators.
In digital geometry processing, a fundamental research task involves partitioning a 2-manifold surface into curved Voronoi cells based on geodesic distances. In this scenario, the 2-manifold surface serves as the domain, the geodesic distance functions as the metric, and the user-specified point set acts as the generators.
A recent advancement by Xin et al.~\cite{xin2022surfacevoronoi} introduces an extensible approach known as SurfaceVoronoi for computing Voronoi diagrams on surfaces and their variants. SurfaceVoronoi operates under the assumption that mesh triangles are small in size and that the triangle-wide geodesic distance field, provided by a single generator, can be considered linear. Leveraging this assumption, SurfaceVoronoi enables each generator to simultaneously propagate distances until all contributing generators for each triangle are identified. Ultimately, for each triangle, SurfaceVoronoi calculates the surface-restricted Voronoi structure by elevating each 2D linear field to a 3D plane and extracting the lower envelope of a roof-like structure.

% introduces two notable complexities. Firstly, the increase in dimensions makes the 3D Voronoi diagram more structurally intricate compared to surface-restricted Voronoi diagrams, which are inherently 2D. Secondly, surface-patch generators yield curved bisectors, diverging from the conventional scenario where point-type generators partition the space using planar walls.

This paper explores the 3D Voronoi diagram of a set of non-intersecting surface patches. We aim to enhance SurfaceVoronoi to tackle this challenge, recognizing that the extension must contend with the increase in dimensions compared to surface-restricted Voronoi diagrams, which are inherently 2D.
In practical applications, the Voronoi diagram of surface patches or even 3D objects proves significantly useful for understanding how objects interact or relate spatially. For instance, Zhao et al.~\cite{zhao2014indexing} proposed using bisectors between two 3D objects to describe their topological relationships.
In the realm of robotics and autonomous systems, the extended Voronoi diagram facilitates the rapid identification of safe paths to avoid collisions or obstacles between 3D objects~\cite{huang2021path}.

While the resulting distance field from a single surface-patch generator can be arbitrarily complex in the entire \( \mathbb{R}^3 \) space, it can be as simple as a linear function when confined to a small range.
This allows us to encode the distance field within a small tetrahedral cell, contributed by a single generator, using a straightforward quadruple. In its nature, this representation signifies a 4D plane, with the first three dimensions denoting coordinates and the fourth dimension illustrating distance variation.
Initially, we generate the initial 4D volume rooted at a tetrahedral element by sweeping the base tetrahedron along the fourth dimension. Our approach begins with the tetrahedralization of the space of interest. Similar to SurfaceVoronoi~\cite{xin2022surfacevoronoi}, the first stage involves propagating straight-line distances from the generators until no generator can offer a smaller distance for any tetrahedral element. Subsequently, we preserve the surviving generators and their corresponding linear distance fields for each tetrahedron.
The second stage involves decomposing each tetrahedron into sub-domains through a sequence of 4D hyperplane cutting operations. Finally, the lower envelope of the 4D roof-like structure, when projected back into 3D, defines the decomposition configuration of the base tetrahedron.
Given that the surface of a CAD model can be readily decomposed into a collection of simple patches, we innovatively apply the extended Voronoi diagram to compute medial-axis surfaces. Extensive experimental results demonstrate that our medial-axis extraction algorithm significantly outperforms the state-of-the-art in terms of accuracy and noise insensitivity. Furthermore, our algorithm can even be used to compute the offset surface.

Our contributions are three-fold:
\vspace{-2pt}
\begin{itemize}
    \item We extend SurfaceVoronoi to compute the Voronoi diagram of a collection of surface patches, addressing a challenging task in past research.
    \item The fundamental operations are elevated from 3D to 4D, enabling the computation of the extended Voronoi diagram, confined within a tetrahedron, through a sequence of 4D hyperplane cutting.
    \item We innovatively apply the new algorithm to compute medial-axis surfaces and demonstrate its superior performance. Additionally, we discuss more potential application scenarios.
\end{itemize}

% t is reasonable to assume
% that the distance field contributed by a single surface patch changes lin-
% early within a tetrahedral range. 

% the most commonly seen setting 
% SurfaceVoronoi. 
% When sources are represented as points and Euclidean distance is used as the metric, the resulting partitioning is called the Voronoi diagram. 
% The current widely adopted approach to compute Voronoi diagrams involves computing the Delaunay triangulation\cite{Green1978ComputingDT} based on the given points and then performing dualization to obtain the Voronoi diagram.
% However, conventional methods fail to resolve this partitioning when sources are not represented as points or when non-Euclidean distance metrics are used.

% When computing Voronoi diagrams, the partitioning planes are always planar structures.
% However, the complexity of the situation increases as more types of sources and distance metrics are allowed, making the computation process more challenging.
% For instance, the resulting partitioning surface takes the form of a hyperboloid when the sources are represented as lines and the Euclidean distance is used as the metric. 
% Similarly, when the sources are points and the distance metric is defined as the Euclidean distance subtracted by different constants (each constant value fixed for each source), the partitioning surface also takes the form of a hyperboloid.
% This highlights the inherent complexity of computing Voronoi-like partitioning in three-dimensional space when dealing with non-point sources or non-Euclidean distance metrics.

\section{RELATED WORKS}
\subsection{Conventional Voronoi Diagrams}
Suppose that we have a set of generators $\mathcal{S} = \{s_i\}_{i=1}^n$ in a given domain~$\Omega$ that is equipped with a metric function~$\mathbf{D}$, 
the Voronoi diagram involves partitioning~$\Omega$ into regions such that
the generator~$s_i$ dominates a region
\begin{equation}
\{x \in \Omega ~\big |~ \mathbf{D}(s_i,x) \leq \mathbf{D}(s_j, x), j \neq i\}.    
\end{equation}
The most prevalent version assumes that~$\Omega$ represents Euclidean spaces, and the generators are exclusively points. The definition of Voronoi diagrams may vary with domains, metrics, and generator types. Voronoi diagrams, as a fundamental tool, have found widespread applications in computer graphics~\cite{Zong2023P2M, Rui2023GCNO, Rui2022RFEPS, wang2022restricted} and image processing~\cite{laraqui2015images, bilius2020efficient}.

There is a substantial body of literature on the computation of Voronoi diagrams, particularly for computing the Voronoi diagram of point-type generators in Euclidean spaces. The most commonly used methods include the divide-and-conquer scheme~\cite{shamos1975closest}, the incremental construction method~\cite{green1978computing}, and Fortune's sweep line algorithm~\cite{fortune1987sweepline}.
Moreover, it should be noted that the lifting technique transforms the Voronoi diagram problem into a convex hull problem by elevating the computation to a higher-dimensional space~\cite{fortune1995voronoi}. Subsequently, parallelization techniques have been applied to the construction of Voronoi diagrams~\cite{wu2023parvoro, wang2014parallel, reem2012projector}.

Voronoi diagrams find diverse applications in digital geometry processing. One common application involves defining the influence area of a mesh vertex, relying on the principles of Voronoi diagrams.
Choi et al.~\cite{choi2020support} utilized Voronoi diagrams in a novel 3D printing approach. This method, compared to traditional lattice methods, effectively reduces stress concentration and eliminates the need for additional support structures.
Xu et al.~\cite{Rui2023GCNO} leveraged Voronoi diagrams to predict normals for undirected point clouds.
To tackle the challenge of polygon meshes being unsuitable for learning-based applications,  Maruani et al.~\cite{maruani2023voromesh} introduced a unique and differentiable surface representation using Voronoi diagrams. 
% \citet{cheng2018constrained} utilized Voronoi diagram to address constrained texture mapping problem. 
% \citet{ying2015point} employs Voronoi diagrams for conducting three-dimensional (3D) point cluster analysis.
% And \citet{amenta1998new} employed the dual characteristics inherent in the Voronoi Diagram and Delaunay triangulation for surface reconstruction.
% \SQ{can you give more related works (just conventional Voronoi diagrams) here?}
% \PF{fixed, I will double check.}

\subsection{Extended Voronoi Diagrams}
In practical applications, Voronoi diagrams may have variant versions based on specific requirements. For instance, generators may include line segments~\cite{karavelas2004robust, held2009topology,https://doi.org/10.1111/j.1467-8659.2012.03058.x}. Xu et al.~\cite{xu2014polyline} investigated the computation of geodesic Voronoi diagrams on a mesh surface, employing polylines as generators. Zong et al.~\cite{Zong2023P2M} adopted the concept of the Voronoi diagram between triangles and proposed an efficient point-to-mesh distance query algorithm. While Voronoi diagrams with line-segment or surface-patch generators can be reduced to a standard Voronoi diagram by sampling generators into a finite set of points, the results are less accurate, and computation is time-consuming~\cite{amenta2001power}.

Conceptually, geodesic distances drive the Voronoi decomposition of a curved surface, but computing geodesic Voronoi diagrams is generally time-consuming~\cite{meng2023efficient}. The Restricted Voronoi Diagram (RVD)~\cite{zong2023parallel, yan2009isotropic} considers surface-restricted decomposition based on the proximity between surface points. However, the dual of the RVD may not be a manifold triangle mesh. Wang et al.~\cite{wang2022restricted} proposed a set of provably effective strategies for addressing this issue.  
Xin et al.~\cite{xin2022surfacevoronoi} introduced a triangle-based lifting technique, enabling the computation of various surface-based Voronoi diagrams.
This method allows for the use of different metrics to measure the distance between two points, such as exact geodesic distances~\cite{QH2016} or Euclidean distances.
In addition, there are some methods based on vector heat~\cite{https://doi.org/10.1111/cgf.13116} and PDE-based~\cite{https://doi.org/10.1111/cgf.12611} approaches for computing geodesic Voronoi diagrams. These methods are computationally fast but sensitive to the quality of triangulation.

There is a deep link between Voronoi diagrams and medial axis surfaces~\cite{amenta2001power}. 
Most of the existing approaches~\cite{bradshaw2004adaptive, sun2014medial} for computing medial-axis surfaces
require sampling points from the surface and computing a conventional Voronoi diagram 
as part of the initialization process. 
However, the discrete sampling operation can be time-consuming and computationally demanding, especially when the number of sample points is large. Additionally, it may compromise the quality of the resulting medial-axis surfaces or even lead to failure on certain thin-plate models.
 Yan et al.~\cite{yan2018voxel} observed the medial axis of a voxel shape can be accurately approximated by the interior Voronoi diagram of the boundary vertices, referred to as the voxel core. Their approach demonstrates exceptional accuracy in approximating the medial axis of smooth shapes, ensuring topological correctness when given a sufficiently high-resolution voxelization of the shape.
 Wang et al.~\cite{wang2022computing} observed that the surface-restricted power cell of each medial sphere indicates the tangential surface regions in contact, aiding in classifying a medial sphere as being on a medial sheet, a seam, or a junction. To the best of our knowledge, this method represents the state-of-the-art in extracting the medial axis surface of a CAD model.
% The medial-axis surface of a 3D shape comprises all points within the shape that have more than one closest point on the shape’s boundary. This fundamental geometric structure has been widely utilized in the approximation, simplification, and analysis of shapes. 
In this paper, we extend SurfaceVoronoi~\cite{xin2022surfacevoronoi} to address this challenging problem and demonstrate its utilities in computing medial-axis surfaces and offset surfaces of a CAD model.

\section{Methodology}
\subsection{Review on SurfaceVoronoi}
SurfaceVoronoi~\cite{xin2022surfacevoronoi} processes a polygonal surface as its input and operates under the assumption that the geodesic distance field at the triangle level, provided by a single generator, can be treated as linear. The SurfaceVoronoi algorithm primarily consists of two stages: distance over-propagation and incremental half-plane cutting.

\paragraph{Distance over-propagation.} 
Given a set of source points $\mathcal{S} = \{s_i\}_{i=1}^n$ on a triangle mesh surface, the algorithm allows each generator to propagate distances simultaneously while maintaining priorities through a priority queue. Distances propagate across triangles, and each triangle retains a list of surviving generators.

Without loss of generality, let $f=\triangle v_1v_2v_3$ represent one of the triangles of the surface. The distances of $v_1, v_2, v_3$ are initialized to $\infty$. When a new generator propagates its distances to $f$, the decision of whether the generator should be kept in $f$ is based on competition with existing generators. Let $s_1,s_2,\cdots, s_k$ be the surviving generators in $f$, and $s_{k+1}$ be the new generator.
If there exists a surviving generator $s_i$ ($1\leq i\leq k$) such that:

\begin{align}
     \mathbf{D}(s_{k+1}, v_1) &\geq  \mathbf{D}(s_{i}, v_1), \nonumber\\
    \mathbf{D}(s_{k+1}, v_2) &\geq  \mathbf{D}(s_{i}, v_2),\\
    \mathbf{D}(s_{k+1}, v_3) &\geq  \mathbf{D}(s_{i}, v_3),\nonumber
    \label{eq:compare}
\end{align}
% \begin{equation}
%     \mathbf{D}(s_{k+1}, v_1) \geq  \mathbf{D}(s_{i}, v_1), 
%     \mathbf{D}(s_{k+1}, v_2) \geq  \mathbf{D}(s_{i}, v_2),
%     \mathbf{D}(s_{k+1}, v_3) \geq  \mathbf{D}(s_{i}, v_3),
%     \label{eq:compare}
% \end{equation}
then $s_{k+1}$ is labeled as an invalid generator for $f$, preventing its propagation to neighboring faces. At the end of the distance over-propagation stage, each triangle accumulates a list of surviving generators and corresponding distance triples.

\begin{figure}[h]
%\vspace{-3.0mm}
	\centering
\begin{overpic}
[width=0.8\linewidth]{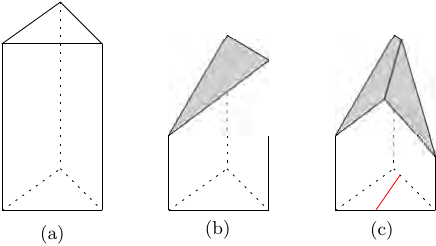}
\end{overpic}
\caption{
% SurfaceVoronoi includes a step of incremental plane cutting: (a)
% The original triangular prism. (b) One cutting operation. (c) Two cutting operations.
% The line segment colored in red reports the triangle-restricted
% Voronoi diagram.
SurfaceVoronoi involves a step of incremental plane cutting, as illustrated in (a) the original triangular prism, (b) after one cutting operation, and (c) after two cutting operations. The red-colored line segment represents the triangle-restricted Voronoi diagram.
}
\label{FIG:planecutting}
% \vspace{-3mm}
\end{figure}
\paragraph{Incremental half-plane cutting.} 
It is straightforward to map~$f=\triangle v_1v_2v_3$ onto a 2D plane,
with the new vertex coordinates being
\begin{equation}
   (x_1,y_1),(x_2,y_2),(x_3,y_3).
\end{equation}
Each surviving generator of~$f$ defines a lifting 3D plane~$\pi$
passing through
\begin{equation}
   (x_1,y_1,d_1),(x_2,y_2,d_2),(x_3,y_3,d_3),
\end{equation}
where $d_1,d_2,d_3$ are the distances from the generator to $v_1,v_2,v_3$, respectively.
\( \pi \) can be implicitly represented as an equation:%\PF{written implicitly as}
\begin{equation}
   d=ax+by+c.
\end{equation}
The lower envelope of the planes established by the surviving generators of $f$ shapes a roof-like structure, using $\triangle v_1v_2v_3$ as the base. Consequently, through a series of plane-cutting operations, the roof-like structure can be defined, revealing the $f$-restricted Voronoi diagram; Refer to Fig.~\ref{FIG:planecutting}.

% \paragraph{Remark.} Extending SurfaceVoronoi to our scenario is not a trivial task. The first major challenge arises from the increase in dimensions.
% The second major challenge is that the use of surface-patch generators results in a curved Voronoi diagram, which is more complex than the conventional situation where point-generator Voronoi diagrams consist of planar bisectors.

\subsection{Linear Representation of Scalar Field in a Tetrahedron}
\label{sec:Alge}
Similar to SurfaceVoronoi, we assume that the tetrahedron-range distance field for a single source~$s_i$ (potentially a surface patch in this paper) undergoes linear changes.
Let $t$ be a tetrahedron with four vertices 
\begin{equation}
     v_1(x_1,y_1,z_1),
     v_2(x_2,y_2,z_2),
     v_3(x_3,y_3,z_3),
     v_4(x_4,y_4,z_4),
\end{equation}
and the distance values given by the generator $s_i$ be $d_1^i, d_2^i, d_3^i, d_4^i$, respectively.
The linear change restricted in~$t$ can be characterized by an equation
\begin{equation}\label{eq:linearDependence}
     d = a_i x + b_i y + c_i z + w_i,
     % \label{equ: hyperplane}
\end{equation}
where $a_i,b_i,c_i,w_i$ can be found by solving
\begin{equation}
\begin{pmatrix}    
   x_{v_1} & y_{v_1} & z_{v_1} & 1 \\
   x_{v_2} & y_{v_2} & z_{v_2} & 1 \\
   x_{v_3} & y_{v_3} & z_{v_3} & 1 \\
   x_{v_4} & y_{v_4} & z_{v_4} & 1\\
\end{pmatrix}
   \begin{pmatrix}
a_i \\
b_i \\
c_i \\
w_i \\
\end{pmatrix}=
\begin{pmatrix}
    d_1^{i}\\
    d_2^{i}\\
    d_3^{i}\\
    d_4^{i}\\   
\end{pmatrix}.
\label{equ:inverse}
\end{equation}
The coefficient matrix is invertible as long as the tetrahedron is not degenerate, ensuring the existence and uniqueness of the solution.

To this end, Eqn.~(\ref{eq:linearDependence}) defines a hyperplane in 4D, where the first three dimensions represent coordinates, and the fourth dimension illustrates distance variation. We assume that the tetrahedron~$t$ has two surviving generators~$s_i$ and~$s_j$. The intersection of their corresponding hyperplanes can be represented as follows:
\begin{equation}
\begin{cases}
a_ix+b_iy+c_iz+w_i=d\\
a_jx+b_jy+c_jz+w_j=d.
\end{cases}
% \begin{aligned}
%     a_i*x+b_i*y+c_i*z+w=e_i\\
%     a_j*x+b_j*y+c_j*z+w=e_j
% \end{aligned}
\end{equation}
By subtracting the two equations, $d$ is eliminated:
\begin{equation}
(a_i-a_j)x+(b_i-b_j)y+(c_i-c_j)z+(w_i-w_j)=0,
\end{equation}
which indicates that the intersection between 4D linear fields defines a 3D plane.

% corresponds to the equation form of a plane in three-dimensional space. 
% It represents the intersection of the tetrahedron with the partition plane formed by the points $s_i$ and $s_j$ within the tetrahedron.

\begin{figure}[h]
%\vspace{-3.0mm}
	\centering
\begin{overpic}
% [width=.98\linewidth]{imgs/lifting4.png}
[width=.98\linewidth]{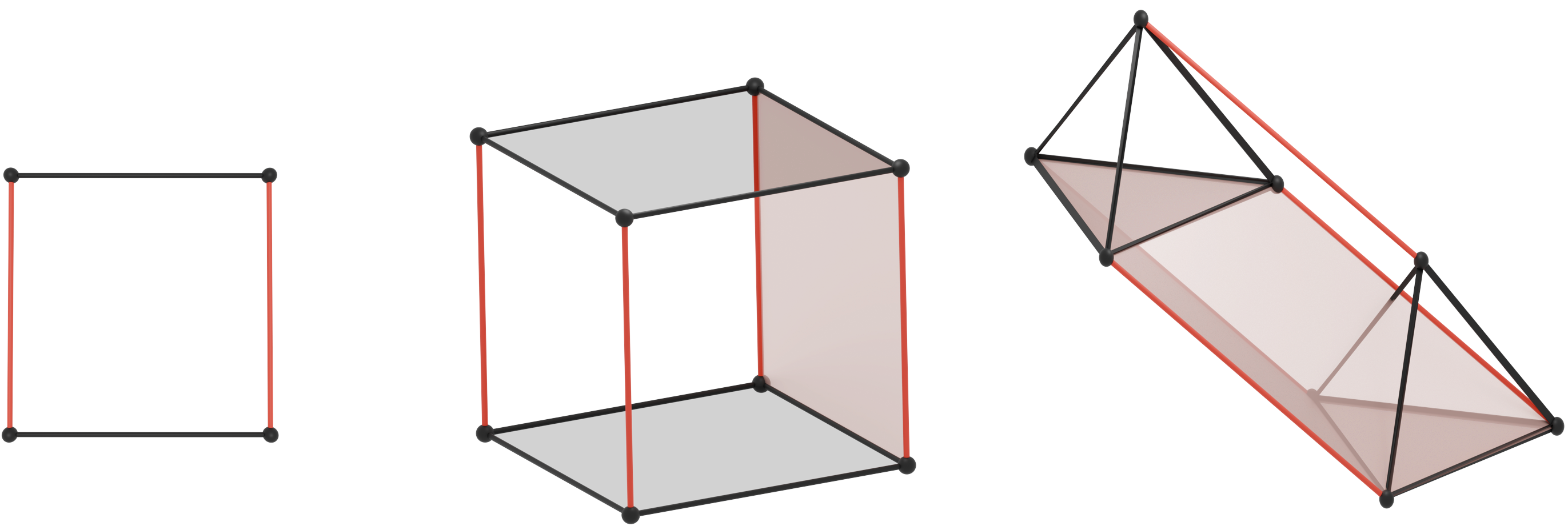}
% \put(19,52){(a)}
% \put(78,52){(b)}
% \put(19,-3){(c)}
% \put(78,-3){(d)}
\put(9,-5){(a)}
\put(45,-5){(b)}
\put(79,-5){(c)}
\end{overpic}
\vspace{3mm}
\caption{
Just as a rectangle (a) and a box (b) can be generated by sweeping a line segment and a rectangle along an additional dimension, respectively, we envision the creation of an initial 4D triangular prism by sweeping a 3D tetrahedron along the fourth dimension (c). In (b) and (c), a 2D side face and a 3D side face are visualized in brown. 
%Figure (d) illustrates the outcome after the shape has been intersected by a hyperplane.
}
\label{FIG:fourDimensionTet}
  \vspace{-2mm}
\end{figure}

\subsection{Geometric View on Tetrahedron-range Linear Scalar Field}
\begin{figure}[h]
%\vspace{-3.0mm}
	\centering
\begin{overpic}
[width=.98\linewidth]{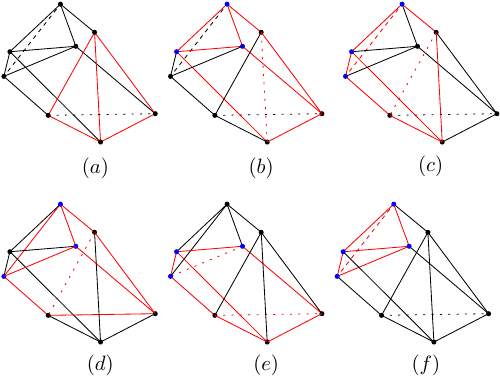}
\end{overpic}
\vspace{3mm}
\caption{
Illustration of how an initial 4D triangular prism is intersected by a 4D hyperplane:  
(a) Base face.  
(b-e) Four 3D side faces generated by sweeping the tetrahedral sides along the fourth dimension.  
(f) Top face.
}
\label{FIG:6sideFace}
  \vspace{-2mm}
\end{figure}

Fig.~\ref{FIG:fourDimensionTet} illustrates the generation of a 4D triangular prism by sweeping a 3D tetrahedron~$t$ along the fourth dimension (extending a triangular prism from 3D to 4D). 
We need to initialize a 4D triangular prism for each 3D tetrahedron.
As $d \geq 0$ in our context, we elevate~$t$ to 4D and consider it as the bottom face of the 4D triangular prism:
\begin{equation}
     (x_1,y_1,z_1,0),
     (x_2,y_2,z_2,0),
     (x_3,y_3,z_3,0),
     (x_4,y_4,z_4,0).
\end{equation}
%\PF{Let \( M \) be set to infinity.}
The top face of the 4D triangular prism is:
\begin{equation}
     (x_1,y_1,z_1,\infty),
     (x_2,y_2,z_2,\infty),
     (x_3,y_3,z_3,\infty),
     (x_4,y_4,z_4,\infty).
\end{equation}
The 4D triangular prism has four vertical edges, respectively connecting 
$(x_i,y_i,z_i,0)$ and $ (x_i,y_i,z_i,\infty)$,
$i=1,2,3,4$. It is bounded by a total of six hyperplanes, 
\emph{i.e.}, the top, the bottom, and four side tetrahedral faces. 
In 4D space, each vertex is formed by the intersection of four hyperplanes. Taking the initial 4D triangular prism as an example, each vertex is created by the intersection of three hyperplanes generated by the prism’s three side faces, along with either the base face or the top face. Similarly, edges in 4D space are formed through a comparable process. We illustrate how an initial 4D triangular prism is intersected by a 4D hyperplane, as shown in Figure~\ref{FIG:6sideFace}.
% Fig.~\ref{FIG:fourDimensionTet}(d) depicts how the initial 4D triangular prism
% is intersected by a 4D hyperplane~$\pi$,
% where the four intersections between~$\pi$ and the vertical edges
% defines the new top face. 

% \begin{figure*}[hb]
% %\vspace{-3.0mm}
% 	\centering
% \begin{overpic}
% [width=\linewidth]{imgs/cut2.eps}
% \end{overpic}
% \caption{
% The 4D triangular prism,
% upon being cut by a 4D hyperplance~(a),
% is bounded by totally six hyperplanes, 
% \emph{i.e.}, the top, the bottom, and four side tetrahedral faces (b-e).
% The blue dots represent three-dimensional entities that currently constitute the four-dimensional structure, points elevated after increasing the dimensionality. 、\PF{I will substitute this diagram.}
% }
% \label{FIG:afterCut}
% \vspace{-3mm}
% \end{figure*}

% For the tetrahedron~$t$, 
% the $t$-restricted linear field can be represented by a quadruple
% that defines a hyperplane~$\pi$ in the 4D space. 
% When~$\pi$ comes,
% we need to first compute the intersections between~$\pi$ and the four vertical edges.
% If~$\pi$ is the first hyperplane to cut the 4D prism,
% the intersection will produce a new tetrahedral top face for the 4D prism; 
% see Fig.~\ref{FIG:afterCut}(a).
% The volume is bounded by totally six hyperplanes, 
% \emph{i.e.}, the top, the bottom, and four side tetrahedral faces; 
% see Fig.~\ref{FIG:afterCut}(b-e).

\section{ALGORITHM}

By considering a set of generators (typically surface patches) in~$\mathbb{R}^3$ as input, we consider computing the Voronoi diagram for these generators. Our algorithm begins with a tetrahedralization of the input, allowing the computation of the complex Voronoi diagram within a tetrahedron range. Subsequently, the Voronoi diagram is calculated through distance over-propagation and incremental hyperplane cutting. The individual steps are detailed in the following subsections.

% needs a tetrahedralized space in which to trace the Voronoi diagram. Like SurfaceVoronoi, it consists of two primary stages, \emph{i.e.},
% distance over-propagation and incremental hyperplane cutting. 

\subsection{Tetrahedralization}
Our algorithm operates within a tetrahedralized space. In general, the computation of the medial axis should be confined to the interior volume enclosed by the input surface. Therefore, the triangles of the original triangle mesh must be embedded into the tetrahedralization results, and each tetrahedron must be sufficiently small. We use fTetWild~\cite{10.1145/3386569.3392385} for this purpose. Additional evaluations regarding how the size of the tetrahedralization affects accuracy can be found in Section~\ref{SEC:CADcalculation}.

\subsection{Incremental Hyperplane Cutting}
\label{Sec:incrementalCutting}
At this point, we assume that the distance between any tetrahedral vertex and any surface patch has been computed, with the distance metric being the Euclidean distance.
This task will be addressed in the next subsection. Following this, we discuss the hyperplane cutting operation within a tetrahedral element.

Given a quadruple~$(x,y,z,d)$, the quadruple is considered to be on the upper side of the hyperplane $$\pi_i:d=a_ix+ b_iy + c_iz + w_i$$ if and only if
\begin{equation*}
    d \geq a_ix+ b_iy + c_iz + w_i.
    \label{equ:upperside}
\end{equation*}
To trace the lower envelope, it is essential to maintain the combination structure, comprising  vertices, edges (connecting two vertices), polygonal faces (formed by three or more edges in a loop), polyhedral faces (having four or more polygonal faces topologically equivalent to a polyhedron), and the 4D convex volume. In our implementation, we must simultaneously retain vertices, edges, polygonal faces, and polyhedral faces.
% When a new hyperplane~$\pi$ comes,
% we first kill the existing vertices that are above~$\pi$,
% as well as edges, polygonal faces and polyhedral faces above~$\pi$,
% through simple tagging.
% We then identify the edges that intersect with~$\pi$ 
% (the edges of the bottom tetrahedron 
%can be disregarded).

When a new hyperplane $\pi$ is introduced, the process begins by eliminating existing vertices, edges, polygonal faces, and polyhedral faces above $\pi$ through simple tagging. The next step involves identifying edges that intersect with $\pi$. For the edge $e=v_1v_2$, if $\pi(v_1)=d_{v_1}-(ax_{v_1}+by_{v_1}+cz_{v_1}+w)>0$ and $\pi(v_2)=d_{v_2}-(ax_{v_2}+by_{v_2}+cz_{v_2}+w)<0$, a new vertex $v$ is obtained using the formula:
\begin{equation}
v=\frac{\pi(v_1)}{\pi(v_1)-\pi(v_2)}v_2-\frac{\pi(v_2)}{\pi(v_1)-\pi(v_2)}v_1.
\label{equ:calNewVertex}
\end{equation}
Following this, it is necessary to update the influenced polygonal faces and polyhedral faces intersecting with $\pi$. 
To elaborate further, if a polygonal face intersects with the hyperplane $\pi$ at two points, then connecting these intersections is necessary to form a new edge. Additionally, if $\pi$ intersects a sequence of polygonal faces, connecting these intersections in a circular order results in the formation of a new 2D side face. Following a similar process, 3D side faces can also be computed.
Finally, each 3D side face is determined by two 4D hyperplanes. If the two hyperplanes that define a 3D side face do not belong to the initial six hyperplanes, the resulting side face contributes to the lower envelope. The lower envelope is then projected back onto the bottom tetrahedron by eliminating the fourth dimension, resulting in a Voronoi-like structure. It is also worth noting that the final structure within a tetrahedral element consists of polygons.

% Finally, the lower envelope is projected back to the bottom tetrahedron by simply eliminating the fourth dimension, resulting in the Voronoi-like structure.

%  \begin{algorithm}[t]
%  \SetAlgoNoLine
%  \KwIn{Tetrahedral $Tet~v_1v_2v_3v_4$ and a set of half-planes~$\{\pi_i\}_{i=1}^k$.}
%  \KwOut{Partition of $Tet~v_1v_2v_3v_4$ based on the lower envelope of $\{\pi_i\}_{i=1}^k$.}
%     Use $Tet~v_1v_2v_3v_4$ to construct a 4D geometric shape with infinite height.\\
%  \ForEach {$\pi\in \{\pi_i\}_{i=1}^k$}
%  {
% Screen out vertices above $\pi$\;
% Screen out useless edges\;
% Screen out useless faces\;
% \ForEach {face $f$ crossing $\pi$}
%  {
%  \ForEach {edge $e$ in $f$}
%  {
%  \If{edge $e$ corssing $\pi$}{
%  Generate a new vertex\;
%  Update $e$'s endpoint\;
%  }
%  }
%  Generate a new line segment using the newly generated points and update the structure of $f$\;
%  }
% Generate a new face using the newly generated line segments.\;
%  }
%  Output the alive faces as the VD in $Tet~v_1v_2v_3v_4$\;
%  \caption{Incremental hyperplane cutting}
%  \label{alg:cutting}
%  \end{algorithm}

Furthermore, although several computational geometry libraries (e.g., CGAL~\cite{cgal:hhkps-nt-23a}) provide robust geometric predicates and kernels, indiscriminately replacing double with exact data types may incur substantial algorithmic inefficiencies.
Thus, we need to develop robust algorithmic implementations.

\paragraph{Handling numerical issues.}
When computing the intersection between a hyperplane and an edge with two endpoints $v_1, v_2$, it is crucial to determine the side of each endpoint. According to Eqn.~(\ref{equ:calNewVertex}), if $\pi(v_1)$ and $\pi(v_2)$ are close to 0, there may be a numerical issue. Therefore, we introduce a tolerance $\epsilon$ to measure the degree to which $\pi(v_1)$ and $\pi(v_2)$ are close to 0. We handle the numerical issues by considering the following 8 cases:
\begin{enumerate}
    \item $\big|\pi(v_1)\big|\geq \epsilon$, $\big|\pi(v_2)\big|\geq \epsilon$ and $\pi(v_1)\times\pi(v_2)<0$: 
    Compute the intersection following Eqn.~(\ref{equ:calNewVertex}).
    \item $\pi(v_1)\geq \epsilon$ and $\pi(v_2)\geq \epsilon$: 
    Deem the segment $v_1v_2$ to be above~$\pi$  and discard the segment.
    \item $\pi(v_1)\leq -\epsilon$ and $\pi(v_2)\leq -\epsilon$: 
    Allow the segment $v_1v_2$ to survive.
    \item $\pi(v_1)\geq \epsilon$ and $\big|\pi(v_2)\big|< \epsilon$: 
    Deem the segment $v_1v_2$ to be above~$\pi$  and discard the segment.
 \item $\pi(v_1) \leq -\epsilon$  and $\big|\pi(v_2)\big|< \epsilon$: 
   Compute the intersection following Eqn.~(\ref{equ:calNewVertex}).
    \item $\big|\pi(v_1)\big|< \epsilon$ and $\pi(v_2)\geq \epsilon$: 
     Deem the segment $v_1v_2$ to be  above~$\pi$ and discard the segment.
   \item $\big|\pi(v_1)\big|< \epsilon$ and $\pi(v_2)\leq -\epsilon$: 
  Compute the intersection following Eqn.~(\ref{equ:calNewVertex}).
    \item $\big|\pi(v_1)\big|< \epsilon$ and $\big|\pi(v_2)\big|< \epsilon$: 
     Deem the segment $v_1v_2$ to be above~$\pi$ and discard the segment.
\end{enumerate}
In our experimental setting, the value of $\epsilon$ is set to $10^{-9}$. Based on our tests, most numerical issues can be addressed in this way. Despite its practical usefulness, introducing tolerance may not consistently resolve all numerical issues.
For example, we have identified some rare cases where a hyperplane intersects with existing planes in more than two points. In the event of such occurrences, we need to seek a more robust implementation.
% The aforementioned approximation may result in the planes organized within the topological structure no longer strictly lying in the same plane.
% When a new hyperplane arrives, it may intersect with existing planes in more than two points, leading to topological inaccuracies.
% However, in all examples provided in the paper, such errors did not occur. 
% This phenomenon is only observed in extremely rare cases.
%In case of its occurrence, we need to seek for a more robust implementation. 

\paragraph{Robust implementation.}
Considering that in our algorithm, two key operations include:
\begin{enumerate}
    \item Identifying on which side a point is located with regard to a 4D plane.
    \item Finding the intersection between an edge and a 4D plane.
\end{enumerate}
To overcome possible error accumulation, we encode each intersection~$\mathbf{v}\triangleq (x,y,z,d)^{T}$ with four 4D planes, similar to~\cite{du2022robust}, rather than directly using the approximate coordinates. Each of the four 4D planes has an implicit form:
\begin{equation}
    \pi_i : \mathbf{g}_i^T\begin{pmatrix}
x\\ 
y\\ 
z\\
d
\end{pmatrix}+w_i=0,\quad i=1,2,3,4.
\end{equation}
To this end, the point~$\mathbf{v}$ satisfies
\begin{equation}
\mathbf{A}\mathbf{v}+\mathbf{w}=\mathbf{0},
\label{equ:line}
\end{equation}
where $\mathbf{A}=\{\mathbf{g}_1,\mathbf{g}_2,\mathbf{g}_3,\mathbf{g}_4\}^T$ and $\mathbf{w}=\{w_1,w_2,w_3,w_4\}^T$.

To determine the side of $v$ with regard to a new plane $\pi$, we need to check whether $\mathbf{g}^T \mathbf{v}+w$ is positive or not, where $\pi$ is defined by $\mathbf{g}$ and $w$. In other words, we need to compute the sign of the following expression:
\begin{equation}
\mathbf{g}^T \mathbf{A}^{-1}(-\mathbf{w})+w.
\end{equation}
If the value of the above expression is greater than or equal to zero, a new vertex will be generated. Unlike the approximate incremental cutting method, we do not explicitly compute the positions of new vertices (i.e., intersections between line segments and hyperplanes). Instead, we encode the new vertex by the four hyperplanes that define it. This technique avoids the need to explicitly compute intersection points.
We also recommend using an exact numerical data type (e.g., Gmpq in CGAL) to ensure computational accuracy. It is important to note that using an exact data type to directly compute intersections can result in increasingly complex rational representations, which may significantly raise computational costs. However, encoding each intersection using four hyperplanes effectively mitigates this issue.

% Although an exact data type is employed, the additional computational
% cost is quite limited. It’s worth noting that using an exact
% data type to compute intersections may result in more and more
% complex rational representations, as well as larger computational
% costs.

% It's worth noting that using an exact data type to compute intersections can lead to increasingly complex rational representations, as well as higher computational costs.
% %It can be seen that the representation mentioned above avoids the need to find the actual intersection points. 
% % \change{It is worth noting that if the value of the above expression is 0, we also consider the sign of the expression to be positive.}
% % We also recommend using an exact numerical data type (e.g., gmpq in CGAL) to ensure accurate computations. Although an exact data type is employed, the additional computational cost is quite limited. 

\subsection{Distance Field Propagation}
\label{DisRieldPropagation} 
In SurfaceVoronoi, a priority queue is utilized to maintain the morphology of wavefronts propagating from near to far, allowing for the simultaneous handling of distances between different generators and triangles. In this process, the computation of the geodesic distance and the distance propagation are coupled. In our scenario, however, the computation of the medial axis involves only Euclidean distances. Therefore, we adopt a different strategy for inferring distances. It requires the following steps to finish the distance computation.

{\bf Initialization.} 
We create a BVH structure for the entire surface \( \mathcal{S} \). Additionally, we create a BVH structure for each of its constituent patches \( {\gamma_i} \), enabling efficient distance queries. Each patch \( \gamma_i \) is composed of triangles, with the surface \( \mathcal{S} \) being the union of all such patches \( \{\gamma_i\} \).
After that, with the support of the BVH of $\mathcal{S}$, we find the closest point $v'$ belonging to $\mathcal{S}$
for each tetrahedral vertex $v$. At the same time, we record the surface patch that accommodates $v'$.

{\bf Distance query.} For each tetrahedron with four vertices $v_1$,$v_2$,$v_3$,$v_4$, we perform the following operations:
\begin{enumerate}
    \item [Step 1.] Initialize a standard queue $\mathcal{Q}$ to accommodate $v_1,v_2,v_3,v_4$, 
    and the surface patch set $\Gamma$ to include the surface patches that provide distances to $v_1,v_2,v_3,v_4$.
    \item [Step 2.] Take out the top vertex $v$ in $\mathcal{Q}$ and 
    query the minimum distance from $v$ to $\mathcal{S}$ using the corresponding BVH. 
    If $v$'s nearest surface patch, say, $\gamma'$, does not belong to $\Gamma$,
    perform hyperplane cutting using the 4D plane defined by $\gamma'$. 
    Push all newly generated vertices to~$\mathcal{Q}$.
    \item [Step 3.] If $\mathcal{Q}$ is empty, the process terminates; otherwise, go to Step~2.
\end{enumerate}

Obviously, the above algorithm does not depend on the order in which the tetrahedra are visited. Therefore, it naturally lends itself to a parallel implementation.

\subsection{Error Analysis}
In this section, we provide an upper bound for the error introduced by the linear approximation of a distance field within a tetrahedron.

We denote any point inside the tetrahedron as \(\mathbf{x}\), with \(d(\mathbf{x})\) representing the true distance value, and \(\tilde{d}(\mathbf{x})\) representing the distance value obtained through linear approximation. They satisfy \(|\nabla d(\mathbf{x})| = 1\) and \( |\tilde{d}(\mathbf{x})| \leq 1 \), respectively. The error function is defined as:
\begin{equation}
e(\mathbf{x}) = d(\mathbf{x}) - \tilde{d}(\mathbf{x}).
\end{equation}
Our goal is to estimate the upper bound of \( |e(\mathbf{x})| \).
The gradient of the error function is given by:
\begin{equation}
\nabla e(\mathbf{x}) = \nabla d(\mathbf{x}) - \nabla \tilde{d}(\mathbf{x}).
\end{equation}
According to the properties of gradients, we have:
\begin{align}
   |\nabla e(\mathbf{x})| &= |\nabla d(\mathbf{x}) - \nabla \tilde{d}(\mathbf{x})| \nonumber\\
   &\leq |\nabla d(\mathbf{x})| + |\nabla \tilde{d}(\mathbf{x})| \\
   &\leq 1 + 1 = 2.\nonumber
\end{align}
Let \(\mathbf{v}_t = \{v_i\}_{i=1}^4\) represent the four vertices of the tetrahedron. Then, the distance error for any vertex within the tetrahedron satisfies
\begin{align}
    |e(\mathbf{x})| ~& \leq~  \min_{v \in \mathbf{v}_t} 2*\|x-v\| \nonumber
\end{align}
Therefore, if we denote the radius of the circumscribed sphere of the tetrahedron by \( h \), the maximum error between the linear distance field and the exact distance field within the tetrahedron is \( 2h \). Thus, in general, smaller and well-shaped tetrahedra yield more accurate linear approximation results. 
However, it is worth noting that the accuracy of the results is independent of the resolution of the triangular mesh. Specifically, when a surface patch is nearly planar, our approach is particularly accurate because the true distance field is linear, which aligns well with our linear approximation.
% Therefore, if we use \(h\) to denote the maximum edge length of the tetrahedron, the maximum variation of the error function \(e(\mathbf{x})\) is:
% \begin{equation}
% |e(\mathbf{x})| \leq 2h.
% \end{equation}
% \PF{Smaller and well-shaped tetrahedra will yield more accurate linear approximation results.}

\section{EVALUATION}
% Our algorithm was implemented using C++ on a platform equipped with a 3.4 GHz AMD Ryzen 9 5950X 16-Core CPU, 64GB of memory, and the Windows 11 operating system.
% Most of the models used in the experiments are taken from the ABC Dataset~\cite{Koch_2019_CVPR}.
Our algorithm was implemented using C++ on a platform equipped with a 3.4 GHz AMD Ryzen 9 5950X 16-Core CPU, 64GB of memory, and the Windows 11 operating system. 
We utilized double precision for data representation and employed the numerical issue handling method described in the paper.
Most of the models used in the experiments are sourced from the ABC Dataset~\cite{Koch_2019_CVPR}.

\subsection{Variant Voronoi Diagrams}
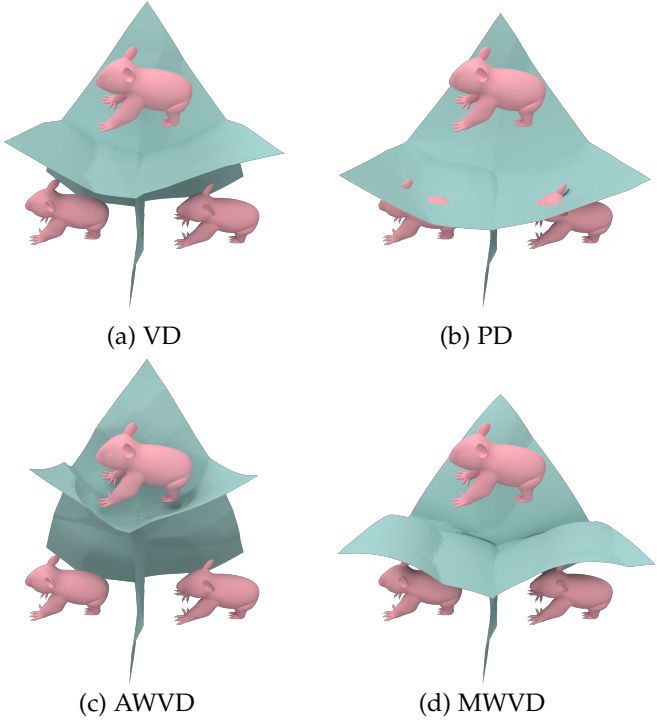
\begin{figure}[h]
%\vspace{-3.0mm}
	\centering
\begin{overpic}
[width=.98\linewidth]{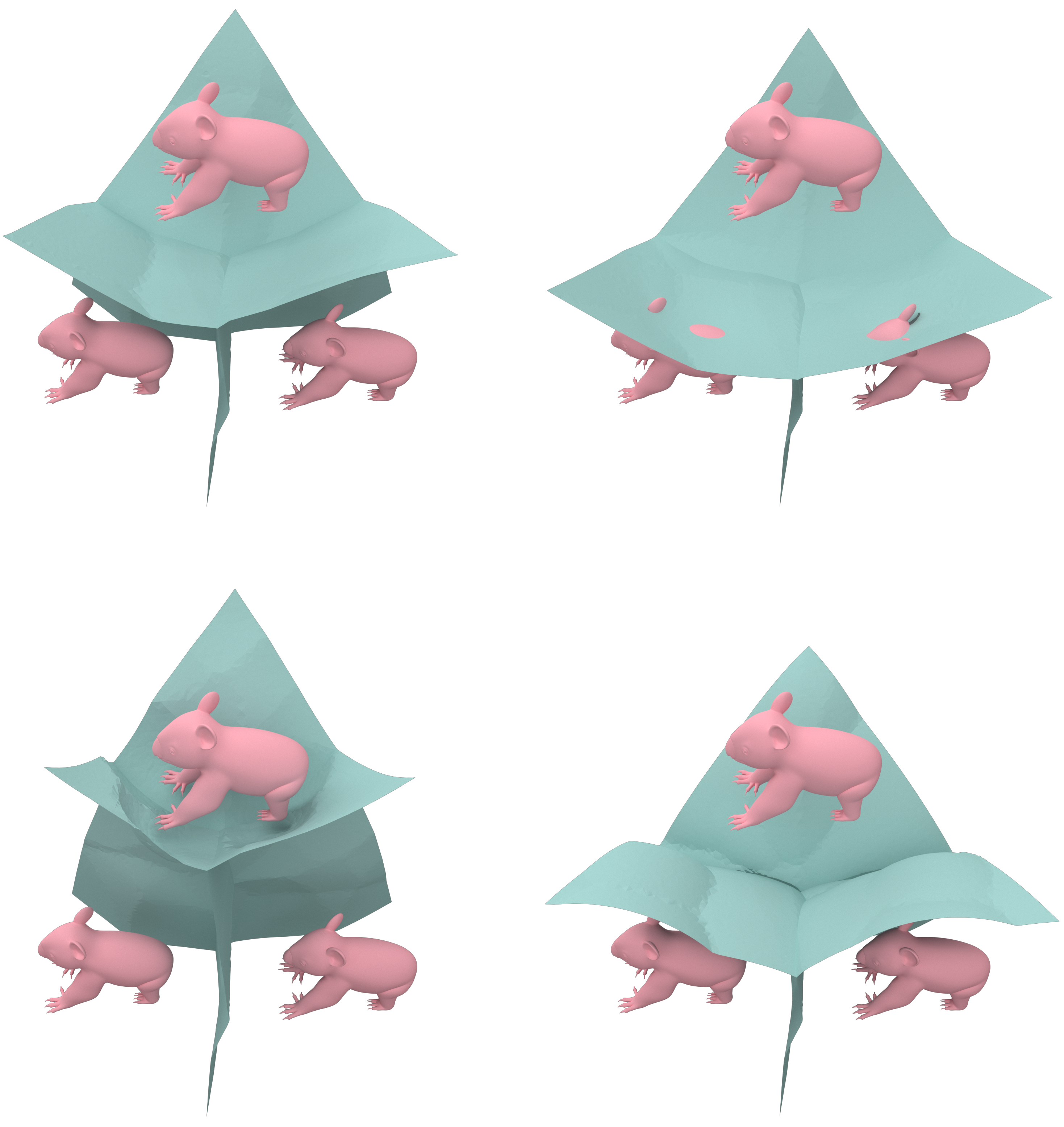}
\put(15,50){(a)~VD}
\put(63,50){(b)~PD}
\put(11,-3){(c)~AWVD}
\put(60,-3){(d)~MWVD}
\end{overpic}
\caption{
Variant versions of Voronoi diagram. 
Here we take four identical Koala models as generators.
%Calculating Voronoi-like diagram using four Koala models as sources.
}
\label{FIG:koala}
  \vspace{-2mm}
\end{figure}

Our method supports the computation of various Voronoi diagram variants. In Fig.~\ref{FIG:koala}, we use four identical Koala models as generators and compute four different Voronoi diagrams, namely:
% In addition to using the Euclidean distance as the distance metric, our method naturally supports other distance measurements.
% In Fig.~\ref{FIG:koala}, we use four identical Koala models as generators and compute four different Voronoi diagrams, namely:
\begin{itemize}
\item Ordinary Voronoi Diagram~(VD),
\item Power Diagram~\cite{imai1985voronoi},
\item Additively Weighted Voronoi Diagram~\cite{Lee1981GeneralizationOV},
\item Multiplicatively Weighted Voronoi Diagram~\cite{ash1986generalized}.
\end{itemize}
In the following, we briefly introduce their definitions except for the most traditional version, \emph{i.e.}, Ordinary Voronoi Diagram.

\textbf{Power Diagram~(PD)} considers both the proximity to points and the influence of weights or powers associated with the points. They have applications in areas such as facility location optimization, where the weight or power of a point represents its importance or capacity. Power diagrams allow for more flexible and customized partitioning of space. The control region of $s_i$ is defined as
$$Cell_{\text{PD}}(s_i)=\{p~|~{\mathbf{D}(s_i,p)}^2-w_i^2\leq {\mathbf{D}(s_j,p)}^2-w_j^2,i\neq j\}.$$

\textbf{Additively Weighted VD~(AWVD)} extends the concept of Voronoi diagrams by assigning weights to represent attributes or values associated with the points. AWVD is useful for spatial interpolation, density estimation, and decision-making problems where the combined influence of multiple factors is considered. The control region of~$s_i$ is defined as
$$Cell_{\text{AWVD}}(s_i)=\{p~|~\mathbf{D}(s_i,p)+w_i\leq \mathbf{D}(s_j,p)+w_j,i\neq j\}.$$

\textbf{Multiplicatively Weighted VD~(MWVD)} incorporates weights as multiplicative factors instead of additive factors. This variation is particularly relevant in applications where the weights represent scaling factors or proportional relationships. MWVD has applications in fields such as computational physics, where the weights may represent physical properties or scaling factors for interactions. The control region Cell of $s_i$ is defined as
$$Cell_{\text{MWVD}}(s_i)=\{p~|~\mathbf{D}(s_i,p)\cdot w_i\leq \mathbf{D}(s_j,p)\cdot w_j,i\neq j\}.$$

\subsection{Medial Axis}
\label{SEC:CADcalculation}
\begin{figure*}[htb]
%\vspace{-3.0mm}
	\centering
\begin{overpic}
[width=.98\linewidth]{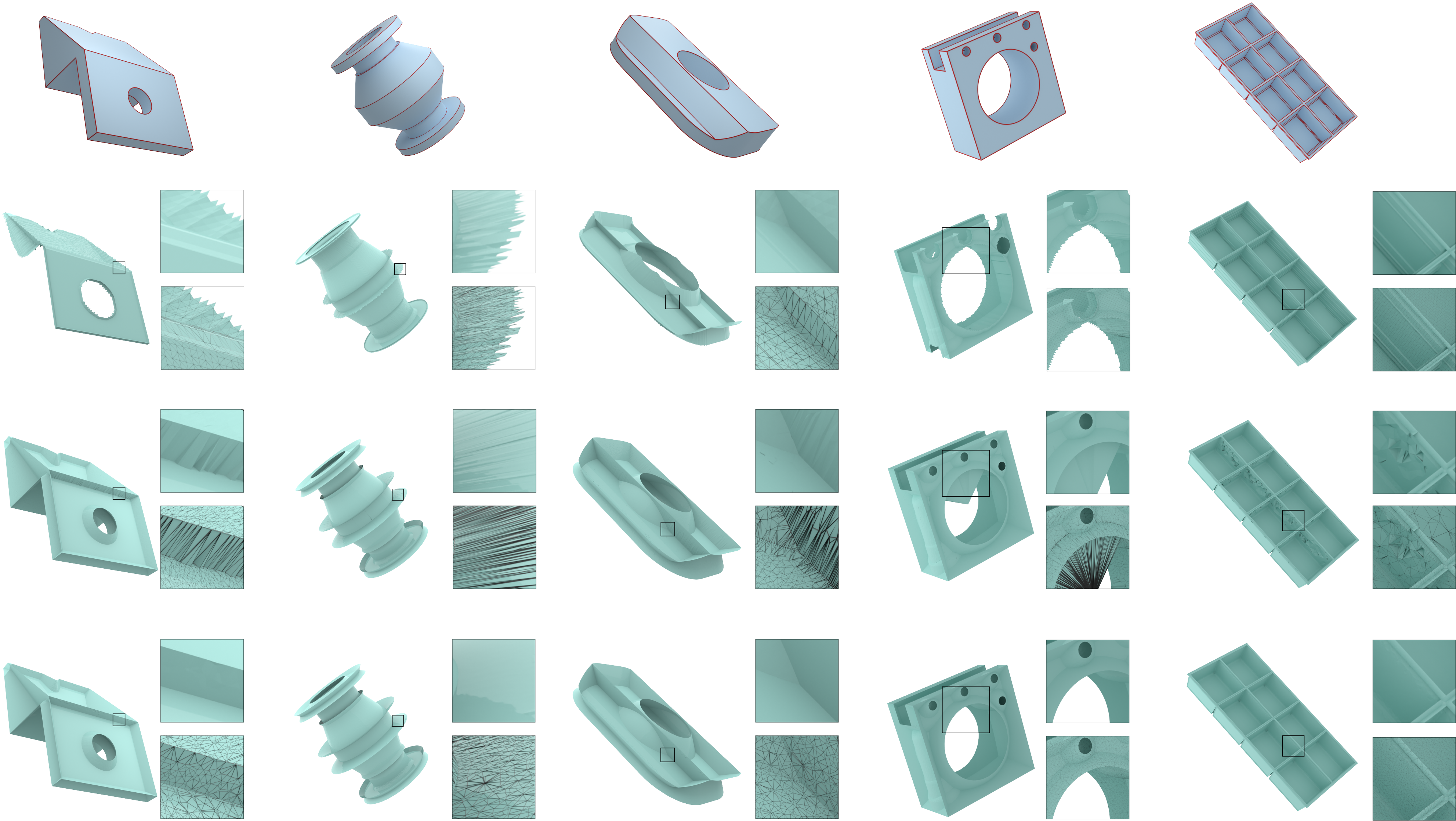}
\put(-2,50){\small{Input}}
\put(-4,36){\small{VoxelCore}}
\put(-4,21){\small{MATFP}}
\put(-3,5){\small{Ours}}

\end{overpic}
\caption{
Visual comparison among Voxel Core~\cite{yan2018voxel}, MATFP~\cite{wang2022computing}, and ours shows that our approach significantly outperforms existing medial axis computation methods on CAD models.
%MATFP results in the presence of zigzag and triangle flipping.
}
\label{FIG:CompareRes}
\end{figure*}
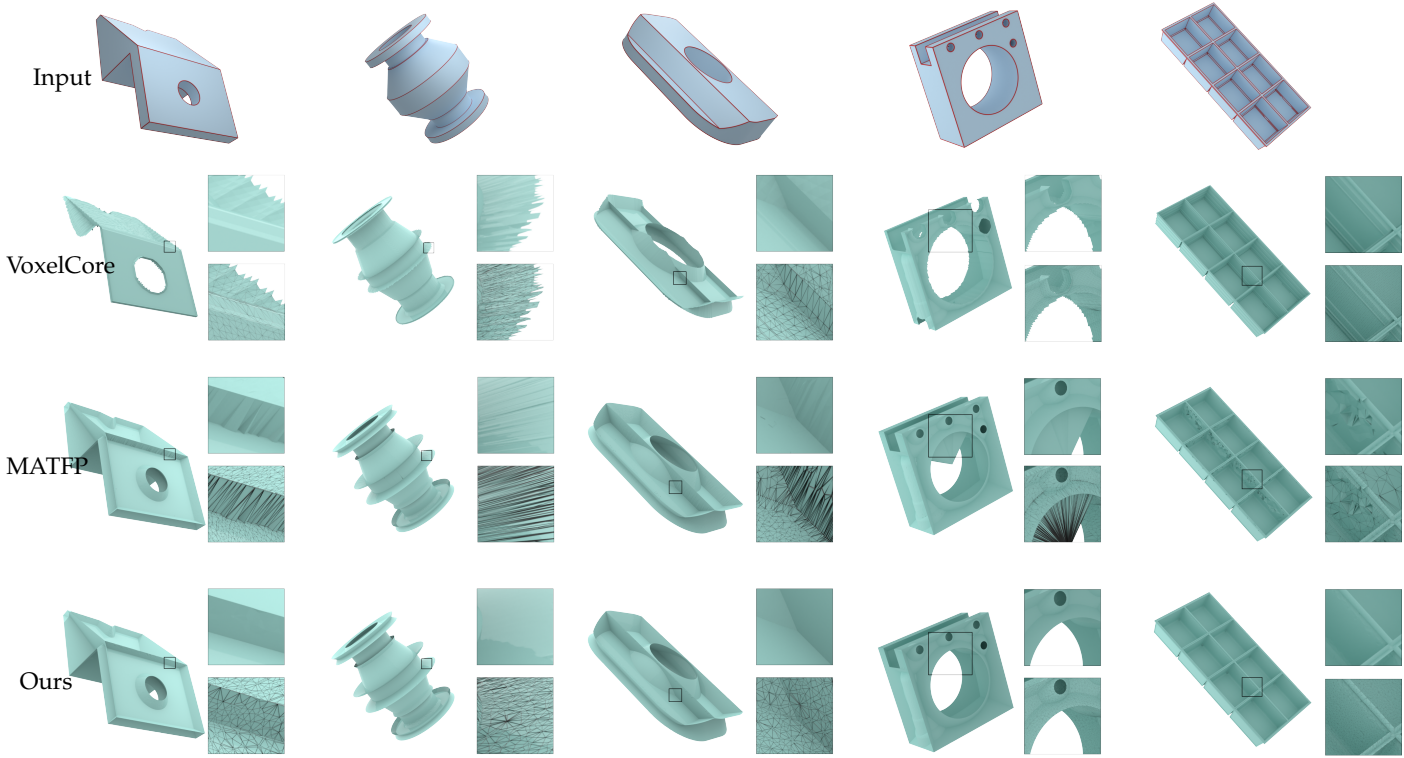

\paragraph{Relevant approaches for comparison.}
Numerous approaches~\cite{yan2018voxel,wang2022computing,li2015q} have been proposed for the medial-axis surfaces problem. The relevant approaches for comparison include:
\begin{enumerate}
    \item VoxelCore: Yan et al.~\cite{yan2018voxel} suggested identifying core voxels (deemed to be located on the medial-axis surface) from all voxels.
    \item MATFP: Wang et al.~\cite{wang2022computing} introduced a method for computing the medial axis of CAD models. MATFP preserves both external and internal features, but the features have to be captured by a seam tracing algorithm. 
\end{enumerate}

% \citet{yan2018voxel} suggested identifying core voxels (deemed to be located on the medial-axis surface) from all voxels.
% Wang et al.~\shortcite{wang2022computing} introduced a method for computing the medial axis of CAD models, named MATFP, which preserves both external and internal features
% but the features have to be captured by a seam tracing algorithm. 

% To ensure a fair comparison, we present the results of MATFP without post-processing.

MATFP requires a step of pre-detecting sharp edges and corners of a CAD model. It initializes with a sampling approach and optimizes their positions before constructing a medial mesh from the updated sphere candidates using restricted regular triangulation. To the best of our knowledge, MATFP represents the state-of-the-art in this field. Similar to MATFP, our approach assumes that the CAD model has been pre-decomposed into multiple surface patches by~\cite{le2017primitive}. Additionally, our algorithm employs fTetWild~\cite{10.1145/3386569.3392385} to tetrahedralize the interior of the input model.

% The method initializes medial spheres using a sampling approach and optimizes their positions before constructing a medial mesh from the updated sphere candidates using restricted regular triangulation. 
% It should be noted that MATFP includes an optional, time-consuming thinning step as post-processing. 

%MATFP requires CAD model with sharp edges and corners pre-detected as the input.
% Similarly, our approach assumes that the CAD model has been pre-decomposed into multiple surface patches~\cite{le2017primitive}.
% Furthermore, we assume that the medial-axis surface is determined by competition between different surface patches, with competition between two points on the same surface patch being disregarded.

% a thinning algorithm after completing all the steps.
% However, in order to compare the inherent performance of the algorithm and considering the longer runtime of the post-processing step, we present the results of MATFP without post-processing. 

\paragraph{Visual comparison.}
Fig.~\ref{FIG:CompareRes} presents a visual comparison among VoxelCore~\cite{yan2018voxel}, MATFP~\cite{wang2022computing}, and our approach.
It can be observed that VoxelCore has at least two disadvantages. Firstly, when the resolution of voxels is insufficient, the accuracy of VoxelCore is significantly reduced. Secondly, the resulting medial axis exhibits many spikes, failing to fully capture the sharp features of CAD models.
The results produced by MATFP successfully capture both the external and internal features of the medial axis. 
Nonetheless, the method involves a step of tracking stitching curves on the medial axis, during which the positions of the sampled points need to be relocated. This can potentially lead to numerous wrinkles on the surface of the medial axis and a significant number of long, thin triangles. In contrast, calculating the medial axis as the Voronoi diagram of patches cleverly avoids this issue.
Additionally, the results calculated by MATFP inevitably contain self-intersections, whereas our results do not exhibit this problem.

\paragraph{Run-time performance.}
Before conducting comparisons, we first analyze the complexity of the algorithm. Consider the scenario of computing the medial axis for a model with \(n\) points within a tetrahedral mesh containing \(n_t\) elements. The time complexity for querying the distance from any arbitrary point in space to the surface of the model is \(O(\log n)\), which is achieved by using a spatial data structure to accelerate distance queries.
Since the volume of each tetrahedron is relatively small, the number of hyperplanes required for incremental cutting within a single tetrahedron remains constant. Thus, the incremental cutting within a single tetrahedron can be completed in \(O(1)\) time.
Therefore, the overall complexity of the incremental cutting process is approximately \(O(n_t \log n)\). For specific models, the runtime of the propagation and incremental cutting of the algorithm increases linearly with the resolution of the tetrahedral mesh. This means that as the tetrahedral mesh resolution increases, the computational time also increases proportionally, but at a linear rate.
 
\begin{table}[]
\begin{center}
\caption{Time statistics (in seconds) of medial axis calculation for the models shown in Fig.~\ref{FIG:CompareRes}.}
\label{Table:TimeStatics}
\setlength{\tabcolsep}{0.1cm} % 调整列间距
\renewcommand{\arraystretch}{1.5} % 调整行间距
\resizebox{1\linewidth}{!}
{
\begin{tabular}{cc|ccccc}
\toprule
\multicolumn{2}{l|}{}                                                                & model 1 & model 2 & model 3 & model 4 & model 5 \\ \hline
\multicolumn{2}{c|}{VoxelCore}                                                              & 1.726   & 1.587   & 0.576   & 2.351   & 1.111   \\ \hline
\multicolumn{1}{c|}{\multirow{2}{*}{MATFP}} & Medial Mesh Initialization                              & 4.153   & 16.877  & 3.588   & 16.980  & 115.272 \\ \cline{2-2}
\multicolumn{1}{c|}{}                       & Thinning                                & 1.537   & 186.675 & 2.010   & 88.692  & 1643.8  \\ \hline
\multicolumn{1}{c|}{\multirow{2}{*}{Ours}}  & fTetwild                               & 12.013  & 31.969  & 13.531  & 43.297  & 28.812  \\ \cline{2-2}
\multicolumn{1}{c|}{}                       & \multicolumn{1}{c|}{Propagation, Cutting} & 2.359   & 4.156   & 1.646   & 5.969   & 5.282   \\ 
\bottomrule
\end{tabular}
}
\end{center}
\vspace{1mm}
\end{table}

\begin{figure}[h]
	\centering
 \hspace{-2mm}
\begin{overpic}
[width=.99\linewidth]{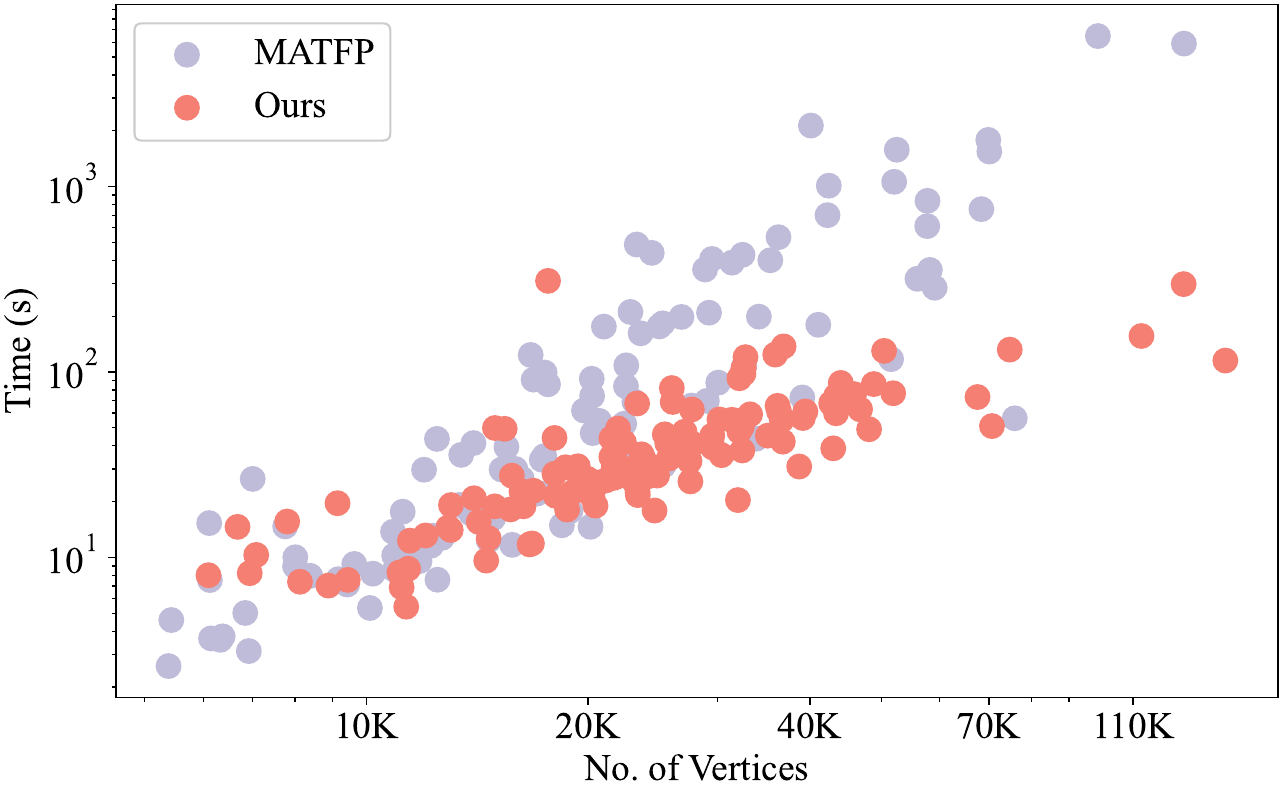}
\end{overpic}
\caption{
Statistics about the computational time on 120 models are provided. MATFP and our approach are marked with grey and red dots, respectively.
}
\label{FIG:26models}
\end{figure}

%-----------------------------------
\begin{figure}[h]
%\vspace{-3.0mm}
	\centering
\begin{overpic}
[width=.99\linewidth]{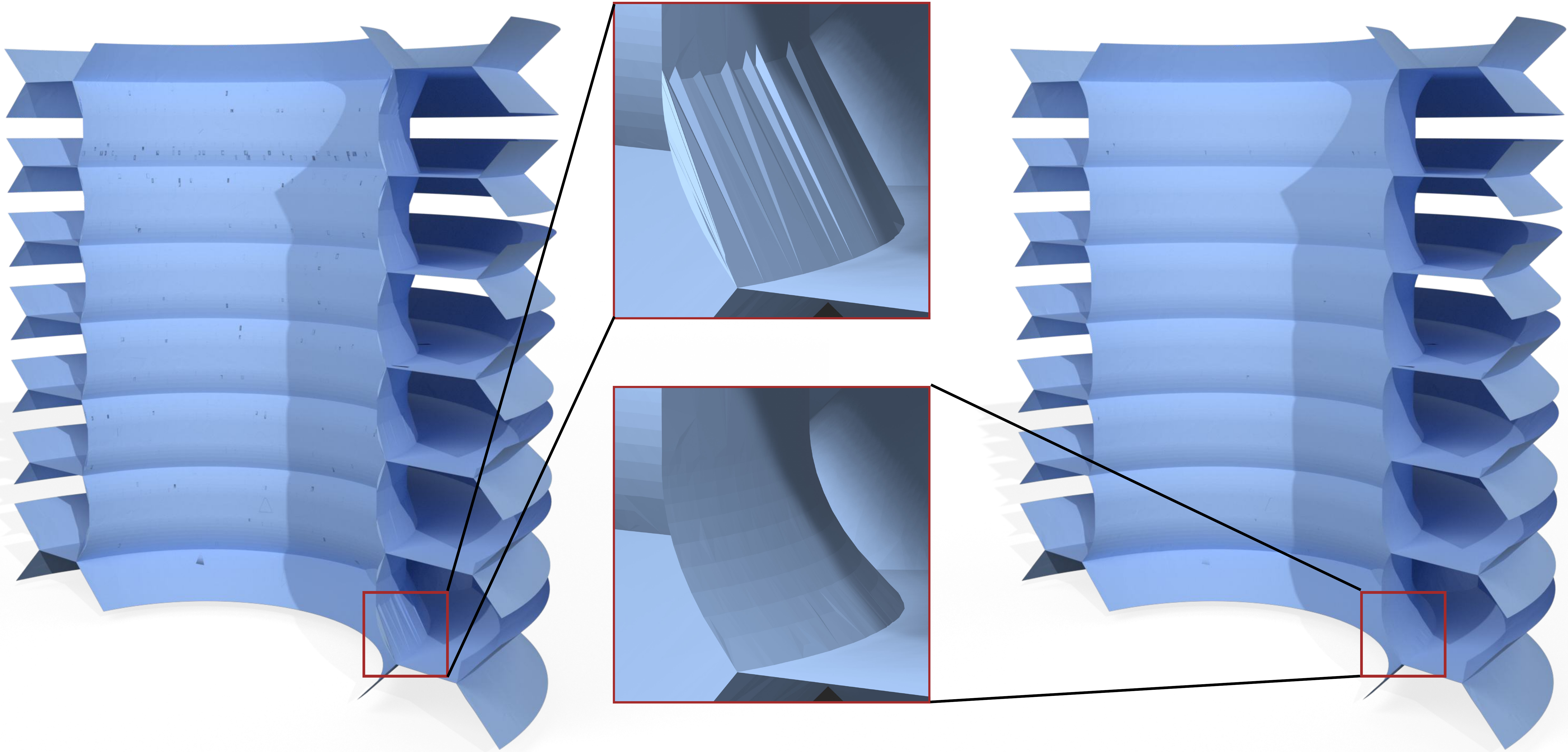}
\end{overpic}
\caption{
The sectional view of the medial axis before and after MATFP's thinning process. For this model, the thinning step takes 6,388 seconds, whereas our method takes only 72 seconds to achieve comparable results, making it faster than MATFP by two orders of magnitude.
}
\label{FIG:thining}
\end{figure}

\begin{figure}[h]
%\vspace{-3.0mm}
	\centering
\begin{overpic}
[width=.98\linewidth]{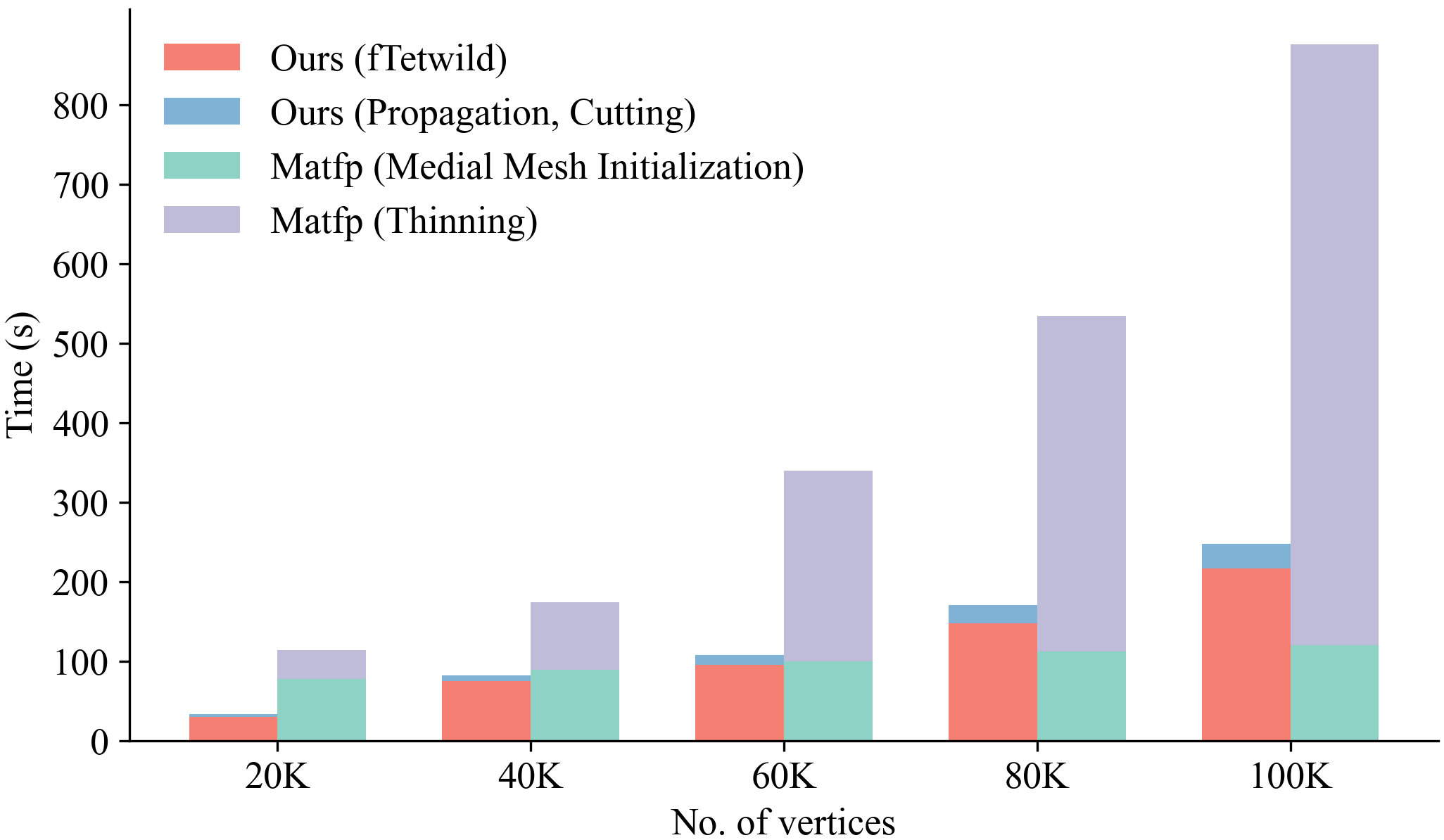}

\end{overpic}
\caption{
The runtime performance statistics of our algorithm indicate that the timing cost incurred in the core steps scales linearly with the resolution of tetrahedralization.
}
\label{FIG:zhuzhuang}
\end{figure}

In our methodology, we employ fTetwild as the tetrahedralization solver, setting the target edge length parameter to $0.015$. Simultaneously, for MATFP, we set the downsampling percentage to $0.1$, and for the Voxel Core method, we adjust the voxel sizes to $128^3$. These settings basically ensure a fair comparison by maintaining similar resolutions across methods. It's important to highlight that, within our algorithm, operations such as distance field propagation and incremental hyperplane cutting can be performed independently for different tetrahedra, allowing for a high degree of parallelization. Timing statistics (in seconds) for VoxelCore, MATFP, and our method are comprehensively presented in Table~\ref{Table:TimeStatics}. It is noteworthy that the thinning process in MATFP can be extremely time-consuming, in some cases taking almost half an hour, which occurs for the fifth model shown in Fig.~\ref{FIG:CompareRes}. 
Additionally, Figure~\ref{FIG:26models} illustrates the computation time statistics for the medial axis across 120 randomly selected models of varying scales, comparing the performance between MATFP and our method, with the x-axis representing the number of vertices in the medial axis results.
In some cases, MATFP's thinning process also requires excessively long overhead. This is because the preliminary medial axis computed using the MATFP method may not consist of a collection of two-dimensional sheets, meaning it does not satisfy the condition of being thin without any solids. Therefore, a thinning step was employed for post-processing to address this. Depending on the model, the thinning step can sometimes be time-consuming. As shown in Figure~\ref{FIG:thining}, the medial axis cross-section results before and after the thinning process are presented. For this particular model, the post-processing step took 6388s.
Furthermore, by randomly selecting 10 models and adjusting parameters to achieve different resolutions for the computed medial axis, Figure~\ref{FIG:zhuzhuang} illustrates the relationship between the time cost and the number of vertices in the medial axis, comparing our method with MATFP. It is evident that our method demonstrates increasing runtime performance advantages as the specified accuracy increases.

% \PF{
% Furthermore, Fig.~\ref{FIG:zhuzhuang} plots how the time cost scales with the number of vertices in the medial axis. It can be seen that ours linearly depends on the number of vertices
%  but the trend of MATFP may with the complexity of the input model. 
% % It is noteworthy that, depending on the complexity of different models, the specific time may vary.
% % is employed to illustrate the relationship between the time cost of our method, Matfp, and the number of vertices in the medial axis.
% }

%Furthermore, Fig.~\ref{FIG:zhuzhuang} is utilized to illustrate the correlation between the timing cost of our method and the tetrahedralization resolution.

% \usepackage{multirow}

% \vspace{-5mm}

\begin{figure}[h]
%\vspace{-3.0mm}
	\centering
\begin{overpic}
[width=.99\linewidth]{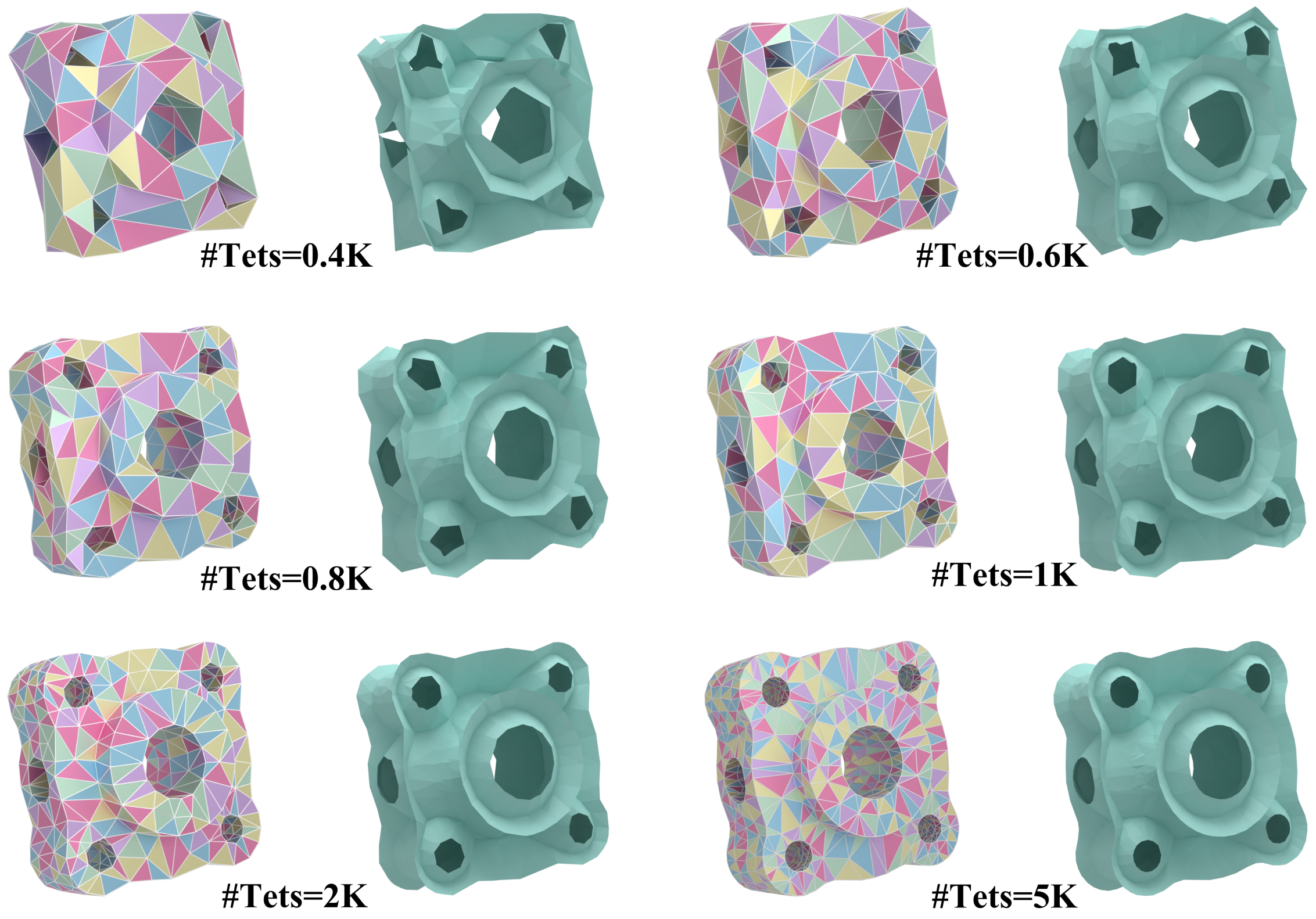}
\end{overpic}
\caption{
We discretize the interior of the input model into different numbers of tetrahedral elements to observe the influence of tetrahedral size. 
We annotated the number of tetrahedra (Tets) in the figure.
% We unify the model and perform tetrahedralization at six different granularities to calculate the model's axis. Each pair on the left side represents the tetrahedralization result of the model, and on the right side is the corresponding axis result. 
%\PF{The model point is randomly moved along the normal direction by the maximum percentage of the diagonal length: $\pm 1.0\%$}
}
\label{FIG:VaryTets}
\end{figure}

\paragraph{Results under different tetrahedral resolutions}
Given our assumption that the distance field within each tetrahedron changes linearly, and considering our practice of performing incremental cutting for each tetrahedron during computation, the size of the tetrahedra plays a crucial role in influencing the outcomes. In Fig.~\ref{FIG:VaryTets}, we employed the fTetwild method on the model at varying tetrahedralization resolutions and showcased the computational findings. It is evident that with an increase in the density of tetrahedralization, the accuracy of the computed results also improves. As demonstrated in Fig.~\ref{FIG:VaryTets}, our methodology is capable of achieving precise results with as few as 2K tetrahedral elements.
% \PF{Because we assume that the distance field within each tetrahedron undergoes linear changes and perform incremental cutting for each tetrahedron during the computation, the size of the tetrahedra has a significant impact on the results. 
% In Fig.~\ref{FIG:VaryTets}, we applied the fTetwild method to the model with varying tetrahedralization resolutions and presented the computational results. 
% It can be observed that, as the tetrahedralization density gradually increases, the computed results become more accurate. 
% In the model depicted in Fig.~\ref{FIG:VaryTets}, with a tetrahedra count of 2k, we can already obtain relatively accurate results.
% }

\begin{figure}[h]
%\vspace{-3.0mm}
	\centering
\begin{overpic}
[width=.98\linewidth]{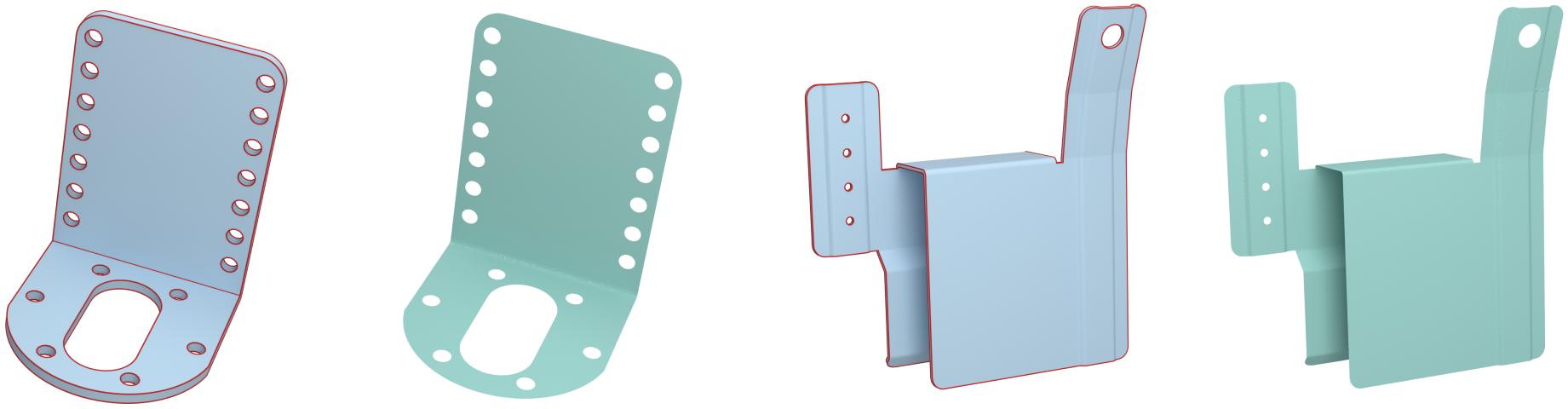}

\end{overpic}
\caption{
Even if the input models are as thin as sheet metal, our algorithm can still generate a faithful medial-axis surface. It's worth noting that, for sheet metals, the side faces are excluded before computing the medial axis.
}
\label{FIG:SheetMetalModel}
  % \vspace{-5mm}
\end{figure}

\paragraph{Sheet metal models}
Our method can be easily extended to compute the medial-axis surface of sheet metal models. The main challenge with sheet metal models is their thinness, which poses a significant hurdle for sampling-based approaches that require a large number of points. However, our approach considers each surface patch as a whole, eliminating the need for a sampling step.
As the side faces typically have minimal impact on the structural behavior during the simulation, they are excluded from the generator list before computing the medial axis.
% As the side faces typically have a small area, they are excluded from the generator list before computing the medial axis. 
% \Q{P7, L32: Why are the side facets excluded from the generator list for sheet metal models?}
Fig.~\ref{FIG:SheetMetalModel} illustrates two typical examples, showcasing the effectiveness of our method.

\begin{figure*}[htb]
%\vspace{-3.0mm}
	\centering
\begin{overpic}
[width=.98\linewidth]{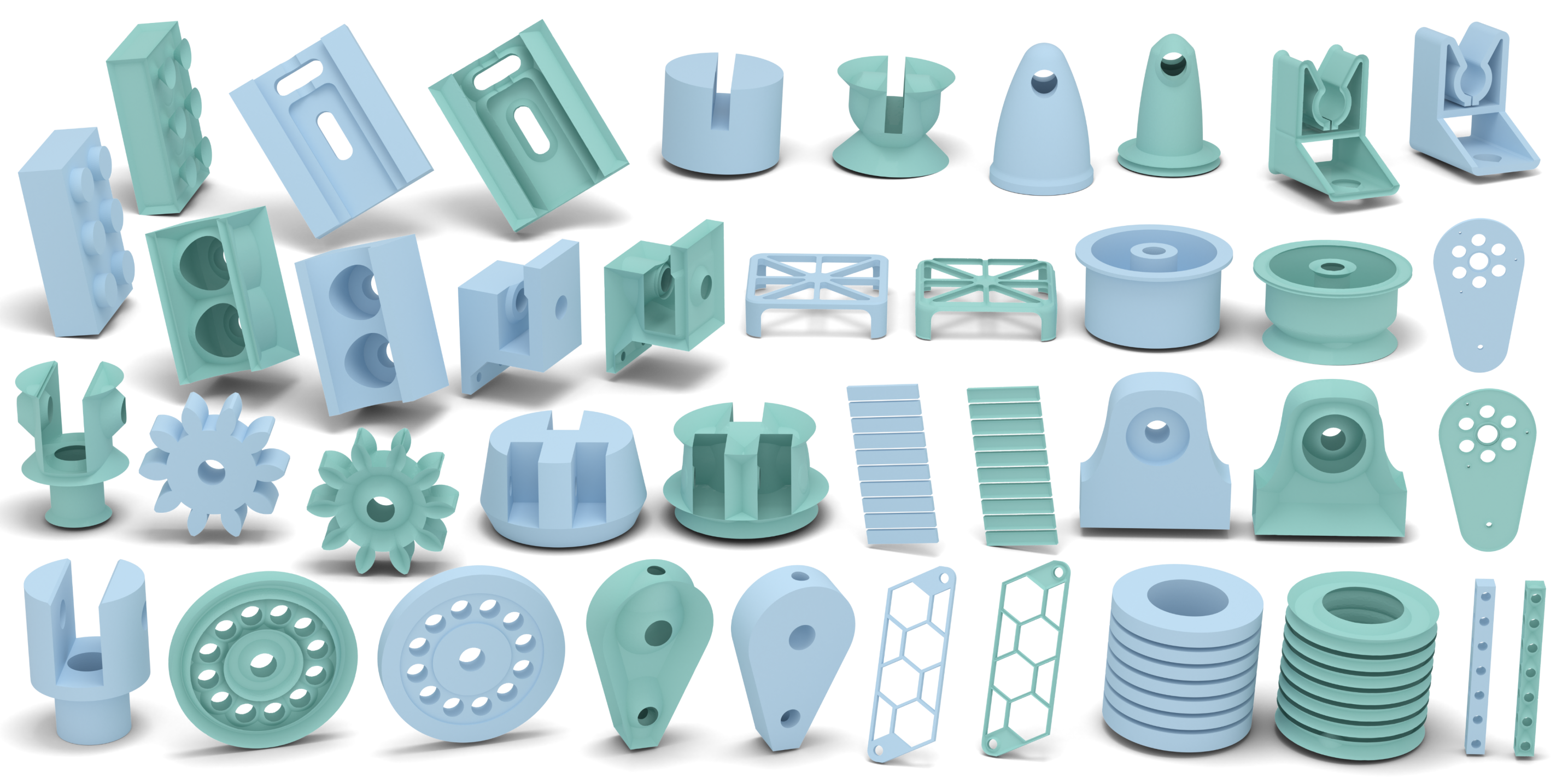}
\end{overpic}
\caption{
More medial-axis results computed by our algorithm.% \PF{canshu qumian}
}
\label{FIG:MedialAxisExample}
\end{figure*}

\paragraph{Organic meshes.}
To validate the effectiveness of the proposed algorithm, we conducted tests on organic meshes. The key difference between organic meshes and CAD models lies in the lack of clear segmentation criteria. We observed that one step of the Variational Shape Approximation method~\cite{10.1145/1015706.1015817} involves partitioning the model’s triangles into several categories. Therefore, we set the total number of categories to 200 and used these categories as surface patches for the algorithm’s input. For organic models, neighboring surface patches typically exhibit smooth transitions, meaning that the Voronoi diagram between them is generally not part of the true medial axis. Therefore, we removed these unnecessary tiny structures during the medial axis computation. As Figure~\ref{FIG:organicmodels} shows, our algorithm still produces fair medial-axis results, demonstrating its great potential.

\begin{figure}[htb]
	\centering
\begin{overpic}
[width=.98\linewidth]{imgs/orgnic.pdf}
\end{overpic}
\caption{
Medial axis computation results on organic meshes.
Different surface patches are represented using distinct colors.
}
\label{FIG:organicmodels}
\end{figure}

\begin{figure}[h]
%\vspace{-3.0mm}
	\centering
\begin{overpic}
[width=.98\linewidth]{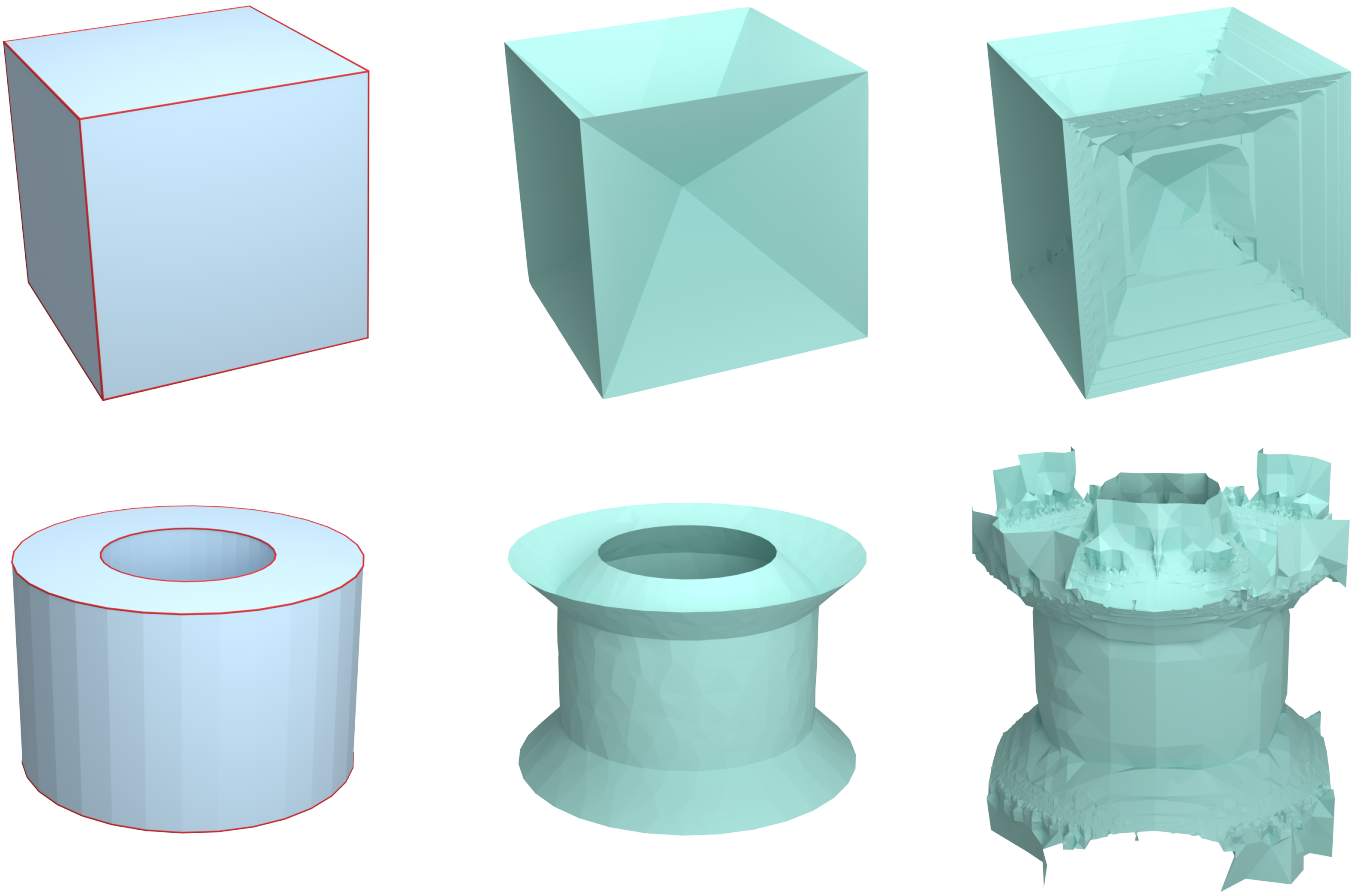}
\put(10,-3){Input}
\put(47,-3){Ours}
\put(69,-3){Edwards et al.~\cite{https://doi.org/10.1111/cgf.12561}}
\end{overpic}
\vspace{3mm}
\caption{
Comparison with Edwards et al.~\cite{https://doi.org/10.1111/cgf.12561}.
}
\label{FIG:gvd}
  % \vspace{-5mm}
\end{figure}
\paragraph{Comparison with generalized Voronoi diagram algorithms.}
There are several methods for computing generalized Voronoi diagrams that take an open or closed surface as a generator. A seminal work in this field was proposed by Edwards et al.~\cite{https://doi.org/10.1111/cgf.12561}. It involves initially constructing an octree, storing essential information at nodes, and subsequently obtaining approximate Voronoi diagram results through reconstruction. 
However, these algorithms face two primary issues when applied to the medial axis computation of CAD models.
On one side, its approximation strategy is not globally accurate, resulting in a lack of smooth transitions at the intersections of different cubes. 
On the other side, the algorithm's configuration of the solution space is not flexible, making it challenging to confine the results within the model's interior.
Figure~\ref{FIG:gvd} demonstrates that the approach by Edwards et al.~\cite{https://doi.org/10.1111/cgf.12561} results in severely bumpy surfaces due to the imprecision of its approximation strategy.

\paragraph{Robustness.}
We randomly selected 500 models from the ABC dataset (all models free of self-intersections) for testing in two modes. This collection encompasses a variety of models including thin plates, slender tubes, and high-genus structures, among others.
When computing the medial axis using the tolerance technique, we encountered 8 instances of numerical errors, resulting in an overall error rate of $1.6\%$. However, by adopting a robust implementation (with exact numerical types discussed in Section~\ref{Sec:incrementalCutting}), we observed no numerical errors.
We provide additional results in Fig.~\ref{FIG:MedialAxisExample}, showcasing a gallery of medial axis surfaces.

\begin{figure}[h]
	\centering
\begin{overpic}
[width=.98\linewidth]{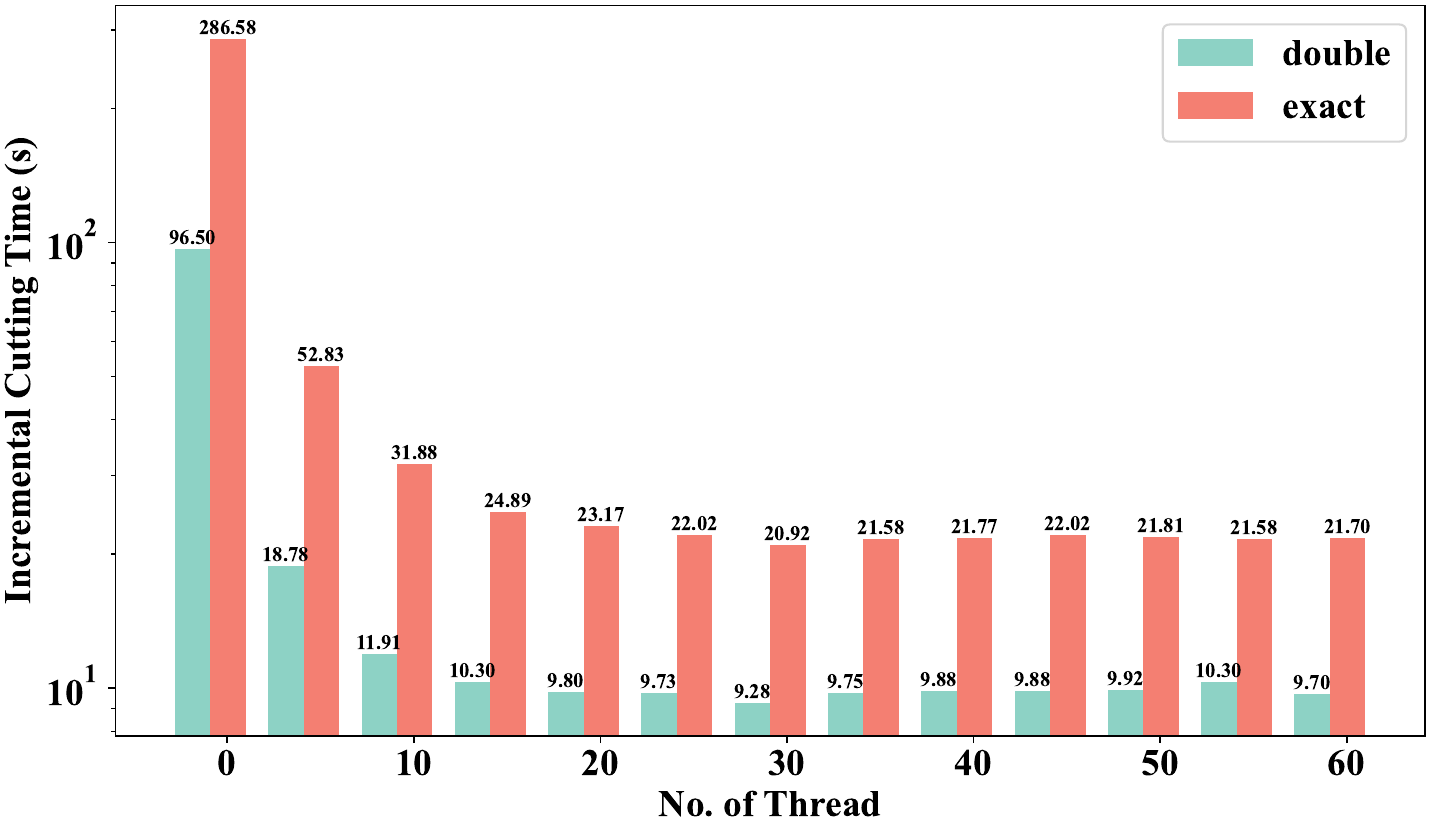}
\end{overpic}
\caption{
Runtime statistics of the exact and approximate incremental cutting strategies under different thread counts for the model in Figure~~\ref{FIG:thining}.
}
\label{FIG:threadtimetest}
\end{figure}
\paragraph{Impact of numeric types and threads on runtime}
In the incremental hyperplane cutting process, the number of threads and the choice of exact/approximate incremental cutting strategies are closely related to the algorithm's runtime. Therefore, we use the model in Figure~\ref{FIG:thining} as an example to study the algorithm's running speed under different thread counts and incremental cutting strategy settings. The results are illustrated in Figure~\ref{FIG:threadtimetest}. It is observed that when the number of threads reaches 32, the speed tends to stabilize, which is also the number of threads chosen in the experiments of this paper. Overall, the runtime of the precise cutting strategy is approximately twice that of the approximate incremental cutting strategy. In addition, we made statistics about the time cost when merely replacing the double data type with an exact type for precise computation, while keeping the number of threads set to 32. The execution time cost in this case is 290.156 seconds. This demonstrates the speed advantage of the precise cutting strategy proposed in this paper.
% \paragraph{More examples.}
%   \PF{Our approach was tested on about 500 models sourced from the ABC dataset, demonstrating that our algorithm consistently delivers highly accurate outcomes. Failures were observed only in cases where the input model contained defects, such as self-intersections.} 

% \PF{Fig.~\ref{FIG:MedialAxisExample} depicts a gallery of medial axis computed by our algorithm. 
% This gallery includes thin plate models, slender tube models,  high-genus models, and so forth. 
% Our algorithm consistently yields highly accurate results across various models.
% \SQ{Figure~\ref{FIG:MedialAxisExample}: say something about 'more examples'}
% \PF{canshu qumian}
% }

\subsection{Offset Surfaces}
\paragraph{Problem statement.}
Offsetting is a fundamental operation in computational geometry and computer graphics. It approximates the shape of a 2D curve or 3D surface by generating a parallel curve or surface at a fixed distance from the original shape. This operation finds various applications, such as creating smooth boundaries around a shape~\cite{22Shell}, performing collision detection~\cite{teschner2005collision}, and generating tool paths for CNC machining~\cite{Kim2007IncompleteMO}.
% offseting is a fundamental operation in computational geometry and computer graphics.
% It approximates the shape of a 2D curve or 3D surface by generating a parallel curve or surface at a fixed distance from the original shape. It has various applications, including creating smooth boundaries around a shape~\cite{22Shell}, performing collision detection~\cite{teschner2005collision}, and generating tool paths for CNC machining~\cite{Kim2007IncompleteMO}.
% \begin{table}[h]
% \resizebox{0.97\linewidth}{!}
% {
% \begin{tabular}{|l|l|l|l|l|l|l|l|}
% \hline
%       & 2005   & 2000:28K & 2003:29k & 2002:30K & 2001:28K & 2008    & 2009   \\ \hline
% MATFP & 14.989 & 21.598   & 26.831   & 68.300   & 31.427   & 109.937 & 79.262 \\ \hline
% Ours  & 6.281  & 15.609   & 10.719   & 22.407   & 17.078   & 7.328   & 11.016 \\ \hline
% \end{tabular}
% }
% \caption{\PF{Time staticais in Fig.~\ref{FIG:CompareRes} (s)}}
% \end{table}
\paragraph{Primary challenge.}
For a given triangle mesh, achieving accurate computation of offsets involves dilating each triangle and explicitly resolving the introduced self-intersections~\cite{jung2004self, campen2010exact, sellan2020opening}. However, these approaches necessitate a tedious post-processing step for handling self-intersections. Alternatively, some methods entail resampling the offset surface by computing the isosurface of the signed-distance function derived from the original surface and then enforcing a deviation~\cite{peternell2007minkowski}. CGAL includes an offsetting function called Alpha Wrapping~\cite{22Shell}.
The resulting output is achieved by greedily refining and carving a 3D Delaunay triangulation on an offset surface of the input, while carving with empty balls of radius alpha.
Nevertheless, these methods may lead to inaccurate offset surfaces with missing sharp features.
The approach proposed by~\cite{https://doi.org/10.1111/cgf.14906} preserves feature information, but its computational process is extremely time-consuming, making it impractical for some applications.

% For a given triangle mesh, accurate computation of offsets can be achieved by dilating each triangle and explicitly resolving the introduced self-intersections~\cite{jung2004self, campen2010exact,sellan2020opening}. However, these approaches have to include a tedious post-processing step of handling self-intersections.
% Alternatively, some methods involve resampling the offset surface by computing the isosurface of the signed-distance function derived from the original surface and then enforcing a deviation~\cite{peternell2007minkowski}. CGAL includes an offsetting function called Alpha Wrapping~\cite{22Shell}. However, these methods may result in inaccurate offset surfaces with missing sharp features.

% Nevertheless, these methods fail to preserve the feature information within the resulting offsets.

\begin{figure}[htb]
%\vspace{-3.0mm}
	\centering
\begin{overpic}
[width=.98\linewidth]{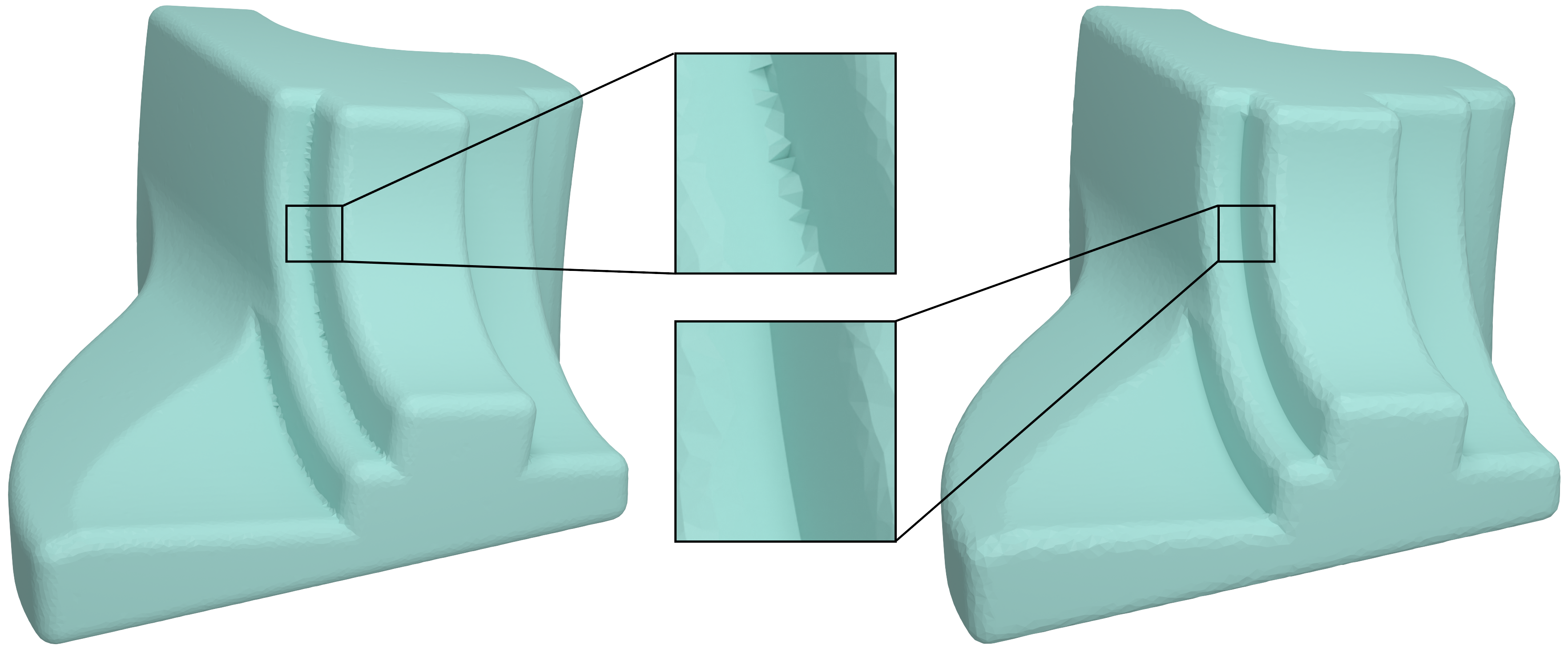}
\put(10,-2){Alpha Wrapping}
\put(80,-2){Ours}
\end{overpic}
\caption{
A visual comparison between Alpha Wrapping~\cite{22Shell} (left) and our method (right) is shown, with close-up windows highlighting the differences. It is evident that our algorithm can preserve sharp feature lines, whereas Alpha Wrapping cannot.
% Visual comparison between Alpha Wrapping~\cite{22Shell} and ours,
% where the close-up windows highlight the differences. 
% It can be seen that our algorithm can retain sharp feature lines but Alpha Wrapping cannot.
}
\label{FIG:offsetBig}
  % \vspace{-2mm}
\end{figure}

\begin{figure}[h]
	\centering
 \hspace{-2mm}
\begin{overpic}
[width=.99\linewidth]{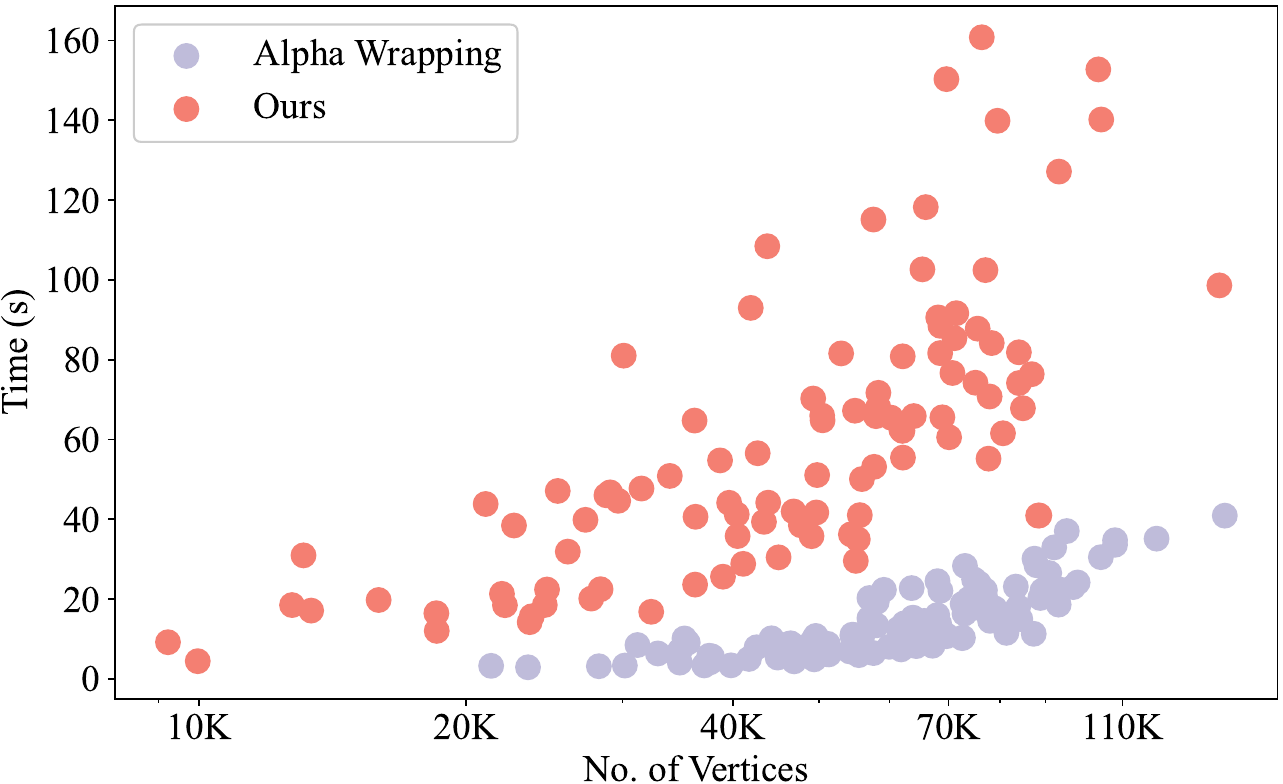}
\end{overpic}
\caption{
Computation time of the offset surface on 100 models of Alpha Wrapping and Ours.
}
\label{FIG:100models}
\end{figure}

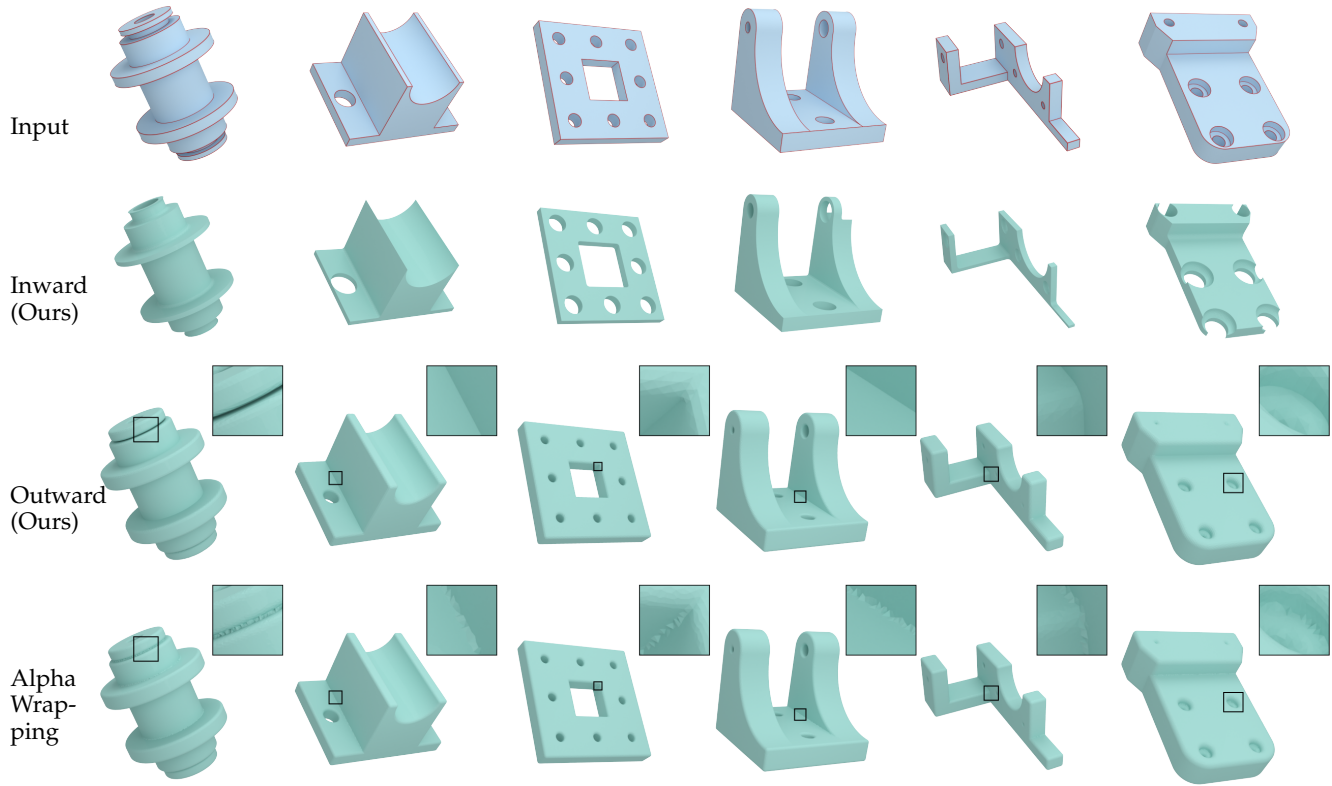
\begin{figure*}[t]
%\vspace{-3.0mm}
	\centering
\begin{overpic}
[width=.9\linewidth]{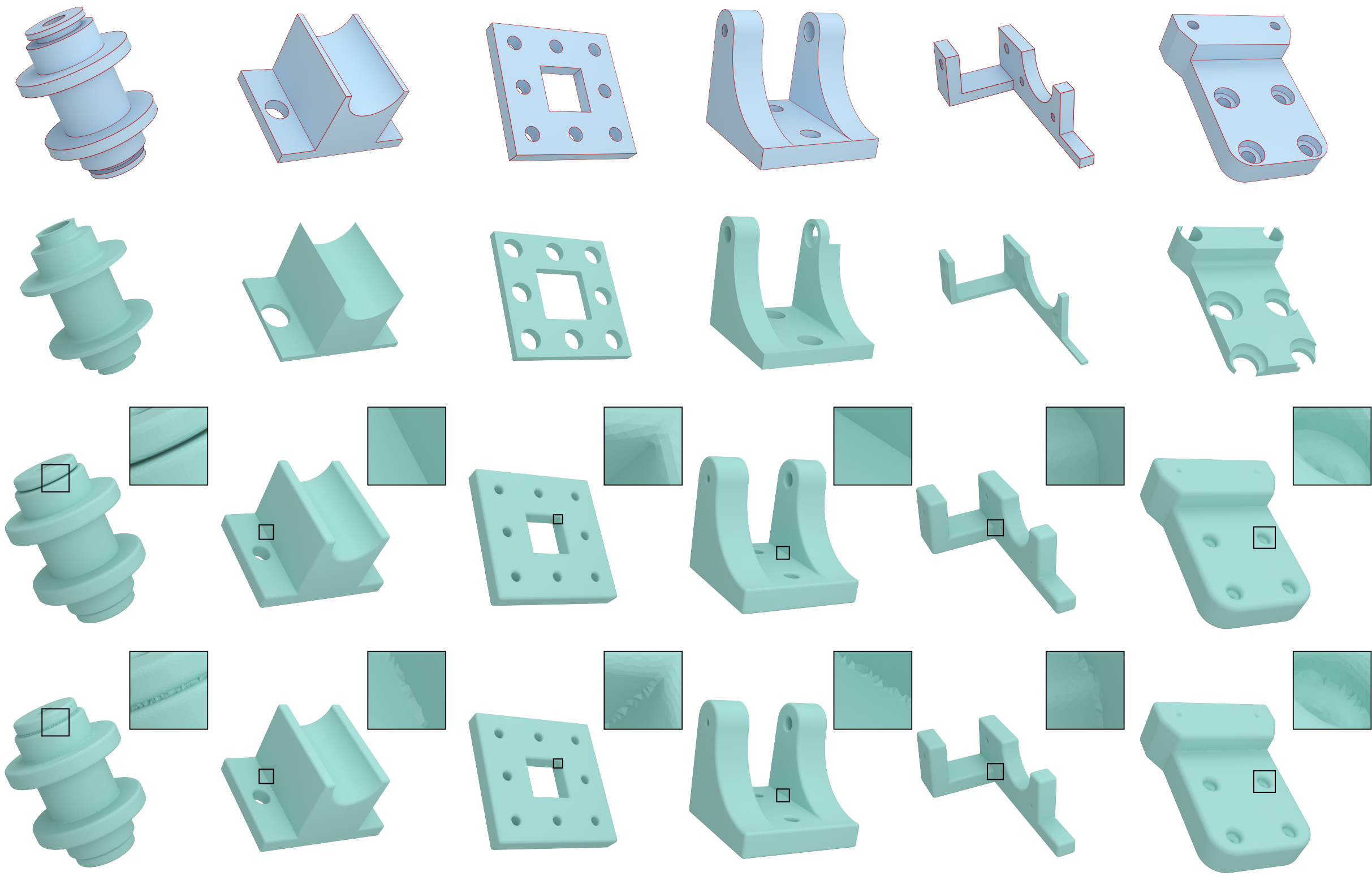}
\put(-7,53){\small{Input}}
\put(-7,39){\small{\parbox{.5in}{Inward (Ours)}}}
\put(-7,22){\small{\parbox{.5in}{Outward (Ours)}}}
\put(-7,6){\small{\parbox{.5in}{Alpha Wrapping}}}
\end{overpic}
\caption{
Visual comparison between Alpha Wrapping~\cite{22Shell} and ours for computing offset surfaces.
% \PF{mesh}
}
\label{FIG:OffsetResComp}
  \vspace{-2mm}
\end{figure*}

% However, computing the offset of a shape presents challenges such as preserving feature information and avoiding self-intersecting structures.
% Our method effectively addresses potential self-intersection issues that may arise during the offsetting process while preserving the feature information of the segmented models in the offset results.
% We provide an implementation of our approach to demonstrate its effectiveness.

% \begin{figure*}[t]
% %\vspace{-3.0mm}
% 	\centering
% \begin{overpic}
% [width=.9\linewidth]{imgs/offset0-4.png}
% \put(-7,53){\small{Input}}
% \put(-7,39){\small{\parbox{.5in}{Inward (Ours)}}}
% \put(-7,22){\small{\parbox{.5in}{Outward (Ours)}}}
% \put(-7,6){\small{\parbox{.5in}{Alpha Wrapping}}}
% \end{overpic}
% \caption{
% Visual comparison between Alpha Wrapping~\cite{22Shell} and ours for computing offset surfaces.
% % \PF{mesh}
% }
% \label{FIG:OffsetResComp}
%   \vspace{-2mm}
% \end{figure*}

\paragraph{Our solution.}
Suppose we are computing an offset surface at a user-specified distance of $d$. Mathematically, we aim to find the the surface where $\big|\mathbf{D}\big|= d$. Considering that a CAD model can be easily decomposed into a collection of simple surface patches, we assume that each surface-patch generator $s_i$ contributes a linear distance field within a tetrahedron $t=v_1v_2v_3v_4$. Therefore, unlike medial-axis extraction, $s_i$ can survive in the tetrahedron $t=v_1v_2v_3v_4$ if:
\begin{enumerate}
    \item $s_i$ is not defeated by other generators in $t$;
    \item $\min_j \{\mathbf{D}(s_i,v_j)\}\leq d\leq \max_j \{\mathbf{D}(s_i,v_j)\}$.
\end{enumerate}
At the end of distance over-propagation, if a tetrahedron contains an empty generator list, it does not contribute to the offsetting surface. 

Our fundamental insight is that the offset surface at distance $d$ can be derived by identifying where the distance field $\mathbf{D}$ matches $2d-\mathbf{D}$. 
It is much like the situation of medial axis computations. For a given tetrahedron $t$, consider its associated surface patch set as $S=\{s_i\}_{i=1}^m$. We then define a corresponding virtual patch set $S^{'}=\{s^{'}_i\}_{i=1}^m$. The distances from $s^{'}_i$ to the vertices $v_i$ are set as $2d-\mathbf{D}(s_i,v_i)$. The distance fields within the set $S$ are used to incrementally cut and preserve the lower envelope within a four-dimensional framework. Concurrently, the distance fields within set $S^{'}$ perform incremental cuts to maintain the upper envelope of the same four-dimensional structure. The intersection of these two envelopes, resulting from the iterative process, indicating the competition between $S$ and $S^{'}$. Through a procedure akin to computing a medial axis, we are thus able to delineate the offset surface, comprising both an inward and an outward layer. Separating these layers based on their connectivity subsequently becomes a straightforward task.

\paragraph{Visual comparison.}
In Fig.~\ref{FIG:offsetBig}, a visual comparison between the Alpha Wrapping algorithm and our method is presented, with differences highlighted in close-up views. The comparison demonstrates that our offsetting algorithm can generate distinctive feature lines, while Alpha Wrapping cannot. Additionally, it is important to note that Alpha Wrapping does not support inward offsetting. More examples for comparison are provided in Fig.~\ref{FIG:OffsetResComp}.

% Fig.~\ref{FIG:offsetBig} presents a comparison between the Alpha Wraping~\cite{22Shell} algorithm and our method,
% where the differences are highlighted in close-up views. 
% It clearly shows that 
% our offsetting algorithm can produce distinctive feature lines,
% but Alpha Wrapping cannot.
% Furthermore, it must be noted that Alpha Wrapping algorithm does not support inward offsetting.
% Fig.~\ref{FIG:OffsetResComp} gives more examples for comparison. 
% \usepackage{multirow}
\begin{table}[]

\begin{center}
\caption{Time statistics (in seconds) of offset calculation for the models shown in Fig.~\ref{FIG:OffsetResComp}.}
\label{Table:offsetTime}
\setlength{\tabcolsep}{0.1cm} % 调整列间距
\renewcommand{\arraystretch}{1.5} % 调整行间距
\resizebox{1\linewidth}{!}
{
\begin{tabular}{cl|cccccl}
\toprule
\multicolumn{2}{l|}{}                                                      & model 1 & model 2 & model 3 & model 4 & model 5 & model 6 \\ \hline
\multicolumn{2}{c|}{Alpha Wrapping}                                        & 24.662  & 12.855  & 7.980   & 11.213  & 6.190   & 11.096  \\ \hline
\multicolumn{1}{c|}{\multirow{2}{*}{Ours}} & \multicolumn{1}{c|}{fTetwild} & 67.613  & 49.344  & 25.375  & 66.829  & 42.516  & 36.547  \\ \cline{2-2}
\multicolumn{1}{c|}{}                      & Propagation, Cutting       & 16.852  & 7.797   & 5.594   & 12.469  & 7.125   & 9.203   \\ 
\bottomrule
\end{tabular}
}
\end{center}

\end{table}

\paragraph{Run-time performance.}
In our experiments, we chose an offset distance equivalent to 2\% of the diagonal length of the model's bounding box. It is evident that the total time required is significantly influenced by the resolution of tetrahedralization; a higher resolution leads to improved accuracy. For the purpose of tetrahedralization, we leverage the space created by offsetting the model's bounding box and embed the input model within the tetrahedralization outcome. Specifically, we set the target edge length for fTetwild to 0.015 and adjust the alpha parameter for the Alpha Wrapping algorithm to 256. The details regarding runtime performance are provided in Table~\ref{Table:offsetTime}.
Additionally, we made statistics about the runtime of the Alpha Wrapping method and ours on 100 randomly selected models, as shown in Figure~\ref{FIG:100models}.
% \PF{
% In our experiment, we set the offset distance to 2\% of the diagonal length of the model's bounding box. 
% Undoubtedly, the total timing cost is highly dependent on the resolution of tetrahedralization, where higher resolution corresponds to higher accuracy. 
% In our approach, we utilize the space obtained by offsetting the model's bounding box as the tetrahedralization space and embed the input model into the tetrahedralization result.
% For the experiment, we configure the ideal edge length of fTetwild as 0.015, and set the alpha value of Alpha Wrapping~\cite{22Shell} algorithm to 256.
% The runtime performance statistics can be found in Table~\ref{Table:offsetTime}.
% }
% report the  in  It is observed that our algorithm generally requires ? seconds to handle a model of approximately ? vertices.

% Therefore
% the offset operation is comparable to that of the core code presented in Table~\ref{Table:TimeStatics}, 
% The segmentation of the model into different surface patches is based on the feature lines present on the input mesh surface.
% The tetrahedral tiling can be computed dynamically during execution, the runtime of the offset operation is comparable to that of the core code presented in Table~\ref{Table:TimeStatics}, given the same resolution.

\paragraph{Results.}
In Fig.~\ref{FIG:koala}, variant Voronoi diagrams are presented, utilizing four identical Koala models as generators. It should be noted that each variant provides a tailored solution for various applications. This example highlights the high flexibility and adaptability offered by our algorithm.

\section{LIMITATIONS AND FUTURE WORK}
Our algorithm, in its current form, exhibits at least three disadvantages. Firstly, it assumes that the tetrahedron-range distance field for a single generator undergoes linear change, which may not hold if the tetrahedral size is large. Secondly, the run-time performance diminishes if the tetrahedralization is too dense. Finally, while our algorithm can produce faithful medial axis and offset results for CAD models, extending it to organic shapes is non-trivial due to the equally challenging problem of decomposing the surface of a free-form shape.

In the future, we plan to address these issues from two perspectives. Firstly, we will introduce adaptive tetrahedralization as an initialization step to balance computational accuracy and run-time performance. Secondly, we will develop improved strategies for decomposing the surface of organic shapes.

\section{CONCLUSION}
In this paper, we extend SurfaceVoronoi to 3D and introduce an accurate algorithm for computing Voronoi diagrams of surface patches. The key observation is that the lower envelope of the 4D roof-like structure encodes the Voronoi diagram structure restricted to a tetrahedron. In implementation, we develop a numerically stable lifting technique for 4D hyperplane cutting operations. The new algorithm operates independently of pre-existing Voronoi numerical packages. We demonstrate its effectiveness in extracting high-quality medial axis transforms (MAT).

Due to its flexibility and scalability, our algorithm can also be adapted to compute the offset surface and various variants of the Voronoi diagram. We provide extensive experimental results to validate its effectiveness and usefulness.

% \bibliographystyle{IEEEtran}

% \bibitem{ref1}
% {\it{Mathematics Into Type}}. American Mathematical Society. [Online]. Available: https://www.ams.org/arc/styleguide/mit-2.pdf

% \bibitem{ref2}
% T. W. Chaundy, P. R. Barrett and C. Batey, {\it{The Printing of Mathematics}}. London, U.K., Oxford Univ. Press, 1954.

% \bibitem{ref3}
% F. Mittelbach and M. Goossens, {\it{The \LaTeX Companion}}, 2nd ed. Boston, MA, USA: Pearson, 2004.

% \bibitem{ref4}
% G. Gr\"atzer, {\it{More Math Into LaTeX}}, New York, NY, USA: Springer, 2007.

% \bibitem{ref5}M. Letourneau and J. W. Sharp, {\it{AMS-StyleGuide-online.pdf,}} American Mathematical Society, Providence, RI, USA, [Online]. Available: http://www.ams.org/arc/styleguide/index.html

% \bibitem{ref6}
% H. Sira-Ramirez, ``On the sliding mode control of nonlinear systems,'' \textit{Syst. Control Lett.}, vol. 19, pp. 303--312, 1992.

% \bibitem{ref7}
% A. Levant, ``Exact differentiation of signals with unbounded higher derivatives,''  in \textit{Proc. 45th IEEE Conf. Decis.
% Control}, San Diego, CA, USA, 2006, pp. 5585--5590. DOI: 10.1109/CDC.2006.377165.

% \bibitem{ref8}
% M. Fliess, C. Join, and H. Sira-Ramirez, ``Non-linear estimation is easy,'' \textit{Int. J. Model., Ident. Control}, vol. 4, no. 1, pp. 12--27, 2008.

% \bibitem{ref9}
% R. Ortega, A. Astolfi, G. Bastin, and H. Rodriguez, ``Stabilization of food-chain systems using a port-controlled Hamiltonian description,'' in \textit{Proc. Amer. Control Conf.}, Chicago, IL, USA,
% 2000, pp. 2245--2249.

% \end{thebibliography}
% \bibliographystyle{plainnat}

\section*{acknowledgments}
The authors would like to thank the anonymous reviewers for their valuable comments and suggestions. This work is supported by National Key R\&D Program of China (2022YFB3303200), National Natural Science Foundation of China (62272277, U23A20312, 62072284, 62172415), the NSF of Shandong Province (ZR2020MF036), the Strategic Priority Research Program of the Chinese Academy of Sciences (XDB0640000 and XDB0640200), and the Beijing Natural Science Foundation (Z240002).

\bibliographystyle{unsrt}

\bibliography{Bibliography}

\vfill

\end{document}